\documentclass[structabstract]{aa}  
\usepackage{natbib}	
\usepackage{ulem}
\usepackage{txfonts,epsfig,graphicx,url,twoopt}
\usepackage[breaklinks=true]{hyperref} 
\hypersetup{colorlinks=true,citecolor=blue}
\bibpunct{(}{)}{;}{a}{}{,}   
\usepackage[usenames,dvipsnames]{color}

\begin{document}

\title{Millimetre spectral indices of transition disks and their relation to the cavity radius}

   \author{P.~Pinilla\inst{1,2}, M.~Benisty\inst{3}, T.~Birnstiel\inst{4}, L.~Ricci\inst{5}, A.~Isella\inst{5}, A.~Natta\inst{6,7}, C.~P.~ Dullemond\inst{2},  L.~H.~Quiroga-Nu\~{n}ez\inst{1}, T.~Henning\inst{8}, L. ~Testi\inst{6,9}}
   \institute{Leiden Observatory, Leiden University, P.O. Box 9513, 2300 RA Leiden, The Netherlands\\
              \email{pinilla@strw.leidenuniv.nl}
        	       \and
	      Universit\"at Heidelberg, Zentrum f\"ur Astronomie, Institut f\"ur Theoretische Astrophysik, Albert-Ueberle-Str. 2, 69120 Heidelberg, Germany
              \and
              Laboratoire d'Astrophysique, Observatoire de Grenoble, CNRS/UJF UMR 5571, 414 rue de la Piscine, BP 53, 38041 Grenoble Cedex 9, France
              \and
              Harvard-Smithsonian Center for Astrophysics, 60 Garden Street, Cambridge, MA 02138, USA
              \and
              California Institute of Technology, MC 249-17, Pasadena, CA, 91125, USA
              \and
              INAF - Osservatorio Astrofisico di Arcetri, Largo Fermi 5, 50125, Firenze, Italy
	      \and
	      School of Cosmic Physics, Dublin Institute for Advanced Studies, 31 Fitzwilliam Place, Dublin 2, Ireland
	      \and
	      Max-Planck-Institut f\"{u}r Astronomie, K\"{o}nigstuhl 17, 69117 Heidelberg, Germany
	      \and
	      European Southern Observatory, Karl-Schwarzschild-Strasse 2, 85748 Garching, Germany}
   \date{Accepted on 24th of February/ 2014}

%%%%%%%%%%%%%
% ABSTRACT
%%%%%%%%%%%%%
 
\abstract
  % context heading (optional)
{Transition disks are protoplanetary disks with inner depleted dust cavities that are excellent candidates for investigating the dust evolution when there is a pressure bump. A pressure bump at the outer edge of the cavity allows  dust grains from the outer regions to stop their rapid inward migration towards the star and to efficiently grow to millimetre sizes. Dynamical interactions with planet(s) have been one of the most exciting theories to explain the clearing of the inner disk.}
% aims heading (mandatory)
{We look for evidence of  millimetre dust particles in transition disks by measuring their spectral index  $\alpha_{\mathrm{mm}}$ with new and available photometric data. We investigate the influence of the size of the dust depleted cavity on the disk integrated millimetre spectral index.}
% methods heading (mandatory)
{We present the 3-millimetre (100~GHz) photometric observations carried out with the Plateau de Bure interferometer of four transition disks: LkH$\alpha$~330, UX~Tau~A, LRLL~31, and LRLL~67.  We used the available values of their fluxes at 345~GHz to calculate their spectral index, as well as the spectral index for a sample of twenty transition disks. We compared the observations with two kinds of models. In the first set of models, we considered coagulation and fragmentation of dust in a disk in which a cavity is formed by a massive planet located at different positions.  The second set of models assumes disks with truncated inner parts at different radii and with power-law dust-size distributions, where the maximum size of grains is calculated considering turbulence as the source of destructive collisions.}
 % results heading (mandatory)
{We show that the integrated spectral index is higher for transition disks (TD) than for regular protoplanetary disks (PD) with mean values of $\bar{\alpha}_{\mathrm{mm}}^{\mathrm{TD}}=2.70\pm0.13$ and $\bar{\alpha}_{\mathrm{mm}}^{\mathrm{PD}}=2.20\pm0.07$ respectively. For transition disks, the probability that the measured spectral index is positively correlated with the cavity radius is 95\%. High angular resolution imaging of transition disks is needed to distinguish between the dust trapping scenario and the truncated disk case.}
% Conclusions
{}
%{Our study shows that high values of the integrated spectral index measured in transition disks do not imply that these disks lack large millimetre grains.}


\authorrunning{P. Pinilla et al.}

\maketitle

%%%%%%%%%%
\section{Introduction}     \label{introduction}
%%%%%%%%%%

Observations by infrared telescopes done with  e.g. Spitzer and Herschel and by millimetre observations with radio-interferometers such as the Submillimetre Array (SMA),  the Very Large Array (VLA) or the Combined Array for Research in Millimeter-wave Astronomy (CARMA) have revealed different types of structures in circumstellar disks surrounding young stars.  Transition disks show a lack of mid-infrared radiation in the spectral energy distribution (SED), implying an absence of warm dust in the inner disk \citep[e.g.][]{calvet2005, pietu06, espaillat10}. Some of these cavities have been resolved by continuum observations \citep[see][for a review]{williams11}.  The recent  discoveries of  companion candidates inside transition disks, such as HD142527 \citep{biller2012} and HD169142 \citep{quanz2013}, promote the idea that holes in transition disks are the result of a brown dwarf, planet, or multiple planets interacting with the disk. However, none of these companion candidates have been confirmed, and alternative explanations have been suggested to  explain the observations \citep{olofsson2013}.
 
Hydrodynamical simulations of planet-disk interactions have been explored for  several years  \citep[e.g.][]{lin79, paardekooper04}. The presence of a massive planet in a circumstellar disk not only affects the surrounding gas, but also the dust distribution in the disk. A planet of $\gtrsim$~1~$M_{\mathrm{Jup}}$ opens a gap in a viscous disk and dust can flow through a gap, depending on the particle size and the shape of the gap \citep{rice06, pinilla12a, zhu2012, zhu2013}. The overall dust distribution is unknown, and observations at different wavelengths with high sensitivity and resolution are crucial for understanding how dust evolves under  disk clearing environments. 

Maps of sub-millimetre and microwave emission from the diffuse interstellar dust have shown that the typical values for the dust opacity index $\beta_{\rm{mm}}$ are $\sim~1.7-2.0$ \citep[see e.g.][]{Finkbeiner1999}. Hence, if the dust in protoplanetary disks is similar to the interstellar medium (ISM) dust, $\beta_{\rm{mm}}$ is expected to be similar. However, changes in the opacity between the ISM and denser regions  have been found \citep[e.g.][]{henning1995}. Dust evolution in protoplanetary disks, and especially the process of grain growth to millimetre sizes in the midplane of disks, has been confirmed by observations \citep[see reviews by][]{natta2007, testi2014}.  The slope of the SED, known as the spectral index $\alpha_{\mathrm{mm}}$, in the millimetre regime can be interpreted in terms of grains size, via  $\alpha_{\mathrm{mm}} \approx \beta_{\rm{mm}} +2$ \citep[see e.g.][]{draine2006},  and low values of  $\alpha_{\mathrm{mm}}$ ($\lesssim 3.5$) correspond to big particles ($\gtrsim$~mm).  Sub- and millimetre observations of classical  protoplanetary disks have shown that the spectral index reaches low values $\alpha_{\mathrm{mm}}\lesssim 3.5$ in Taurus, Ophiuchus, Orion, Chamaeleon, Lupus, etc., star forming regions \citep[e.g.][]{testi2001, Testi2003, rodmann2006, lommen2007, lommen2009, lommen2010, Ricci2010a, Ricci2010b, Ricci2011, ubach2012}. 

Alternative explanations for the low spectral indices in disks have been addressed by different authors. One possible explanation is that the emission or part of the emission may come from unresolved regions where the dust  is optically thick at millimetre wavelengths \citep{testi2001, Testi2003}, but it has been shown that this is unlikely to be the general explanation for the low values of the opacity index measured in large samples of disks \citep{Ricci2012b}. In addition,  the chemical composition \citep{Semenov2003}, porosity \citep{henning1996, stognienko1996}, and  geometry of the dust particles can affect the opacity index. However, their impact is not significant, for reasonable values of the properties of the dust grains  \citep[see e.g.][]{draine2006}. Thus, the most likely explanation is that dust grains have grown to mm-sizes \citep[see in addition][]{calvet2002, natta2004a, natta2007}. 

One of the most critical challenges of  planetesimal formation is to understand how the rapid inward migration of dust grains can stop. The theory of radial drift and observations at mm-wavelength are in strong disagreement \citep[e.g.][]{brauer07}. The confirmation of long-lived presence of mm grains with observations made theoreticians to look for the  missing piece of this rapid inward-drift puzzle. Particle trapping in pressure maxima (i.e.\ locations where $\nabla p=0$ and $\nabla^2p<0$) is a result of the dynamic phenomenon that, due to friction with the gas, particles tend to drift in the direction of higher gas pressure, i.e. up the pressure gradient \citep[e.g.][]{Klahr1997}. Besides the inward drift reduction in pressure bumps, dust grains do not experience the potentially damaging high-velocity collisions due to relative radial and azimuthal drift anymore, thus reducing the fragmentation problem as well. 

Pressure bumps may themselves be caused by planets that have formed by a yet unknown mechanism. A possible first planet embryo can be explained by the growth of a seed (cm-sized object)  by sticking mutual collisions, including mass transfer and bouncing effects as possible outcomes of the collisions \citep{windmark2012}. Once a massive planet is formed and opens a gap, a  pressure bump is expected in the radial direction  at the outer edge of the gap due to the depletion in the gas surface density. This pressure bump becomes a ``particle trap'', a region where particles accumulate and grow. This trap is therefore a good planet incubator for the next generation of planets. Particle trapping creates a ring-like emission at mm-wavelengths, whose radial position and structure depend on the disk viscosity, mass, and location of the planet \citep{pinilla12a}. In addition, the non-axisymmetric emission of transition disks revealed by sub-mm observations \citep{brown09} and recently confirmed with ALMA and CARMA observations \citep{casassus13, vandermarel2013, isella2013, fukagawa2013, perez2014} may also be a result of particle trapping that lead to strong asymmetric maps \citep{birnstiel2013, ataiee2013,  lyra2013}.

Spatially resolved, multi-frequency observations will provide crucial information about the dust density distribution of transition disks. Radial variations in the dust opacity have been revealed for individual disks \citep[e.g.][]{perez2012, trotta2013}; however, a statistical study cannot have been done yet. In this paper, we show that in the case of transition disks,  a change in the cavity radius affects the spatially integrated spectral index, a quantity already available for a significant number of transition disks.
 
We present  3mm-observations of four transition disks: LkH$\alpha$~330, UX~Tau~A, LRLL~31, and LRLL~67  carried out with Plateau de Bure Interferometer (PdBI).  The observations and data reduction of our four targets are presented in Sect.~\ref{observations}, together with the information on additional transition disks. Section~\ref{results} presents the integrated spectral index of our sample with the data collected from the literature. In this section a comparison between the spectral indices of transition and regular protoplanetary disks is presented, as well as a possible correlation between the cavity radii and the spectral index.  In section~\ref{section_discussion} we compare the observations to two different kinds of models: first of all, dust evolution models based on planet-related cavities, where a pressure bump is formed at the outer edge of the gap, which traps millimetre particles; second, models with a power law size distribution of particles with an artificial cut at different radii to imitate dust cavities. Finally, in Sect.~\ref{conclusion} we give the main conclusions of this work.

%%%%%%
\begin{table*}
\label{table1}    
\centering   
\tabcolsep=0.10cm                      
\begin{tabular}{c||cccccccccccccc}       
\\ 
\textbf{{\tiny Name}} & {\tiny Ra} & {\tiny Dec} &  {\tiny SpT} & {\tiny $T_{\mathrm{eff}}$}&{\tiny d}&{\tiny $L_\star$}&{\tiny $R_\star$}&{\tiny $M_\star$}&{\tiny $\dot{M}$}&{\tiny $R_{\mathrm{cav}}$} & {\tiny $F_{\mathrm{0.8mm}}$(mJy)}&{\tiny $F_{\mathrm{3.0mm}} $(mJy)}& {\tiny $F_{\mathrm{1.0mm}} $}&$\alpha_{0.88-3\mathrm{mm}}$\\ 
&{\tiny [J2000]}&{\tiny [J2000]}&&{\tiny (K)}&{\tiny (pc)}&{\tiny ($L_\odot$)}&{\tiny ($R_\odot$)}&{\tiny ($M_\odot$)}&{\tiny ($M_\odot~$yr$^{-1}$)}&{\tiny (AU)}& {\tiny$\pm \sigma$(mJy/beam)} & {\tiny$\pm \sigma$(mJy/beam)} &{\tiny (mJy)} &\\
\hline
\hline
\\
{\tiny LkH$\alpha$~330}&{\tiny03 45 48.29}&{\tiny +32 24 11.9} &{\tiny G3}&{\tiny5830}&{\tiny 250}& {\tiny 15}&{\tiny 3.75}&{\tiny 2.2}&{\tiny 2$.0 \times 10^{-9}$}&{\tiny 68$^{R}$}& {\tiny 210$\pm$2.1}& {\tiny 3.9$\pm$0.12}& {\tiny 138.6$\pm$21.1}& {\tiny 3.25$\pm$0.40}\\ \\ 

{\tiny UX~Tau~A}&{\tiny 04 30 04.00}&{\tiny +18 13 49.3} & {\tiny G8}&{\tiny5520}&{\tiny 140}& {\tiny 3.5}&{\tiny 2.05}&{\tiny 1.5}&{\tiny $1.0\times10^{-8}$}&{\tiny 25$^{R}$}& {\tiny 150$\pm$1.5}& {\tiny 10.1$\pm$0.33}& {\tiny 113.2$\pm$17.2}& {\tiny 2.20$\pm$0.40} \\ \\

{\tiny LRLL~31}&{\tiny 03 44 18.00}&{\tiny +32 04 57.0}&{\tiny G6}&{\tiny5700}&{\tiny 315}& {\tiny 5.0}&{\tiny 2.3}&{\tiny 1.6}&{\tiny $1.4\times10^{-8}$}&{\tiny 14$^{S}$}& {\tiny 62$\pm$6.0}& {\tiny 2.88$\pm$0.08}& {\tiny 45.0$\pm$8.1}& {\tiny 2.50$\pm$0.44} \\ \\

{\tiny LRLL~67} &{\tiny 03 44 38.00} &{\tiny +32 03 29.0} &{\tiny M0.75}&{\tiny3720}&{\tiny 315}& {\tiny 0.5}&{\tiny 1.8}&{\tiny 0.5}&{\tiny $1.0\times 10^{-10}$}&{\tiny 10$^{S}$}& {\tiny 25$\pm$11}& {\tiny 1.21$\pm$0.09}& {\tiny 18.2$\pm$8.5}& {\tiny 2.47$\pm$0.92} \\ \\
\hline
\end{tabular} 
\caption{{\bf Observed targets.} Column~1: Name of the disks observed with PdBI. Column~2 and~3: Right ascension and declination of the targets. Column~4: Spectral type. Column~5: Effective temperature of the central star. Column~6: Estimated distance to the target. Column~7, 8 and 9:  Stellar luminosity, radius and mass respectively. Column~10: Accretion rate. Column~11: Radial cavity extension ($R$ and $S$ implies that the cavity is either resolved or derived from SED modelling, respectively). Column~12: Flux at~345~GHz. The data is from \cite{andrews11}  for LkH$\alpha$~330 and UX~Tau~A, and from \cite{espaillat12} for LRLL~31 and LRLL~67 and references therein. Column~13: Flux at~100~GHz from the PdBI observations. Column~14: Calculated flux at~300GHz ($\sim~1$mm) and the error assuming an additional systematic uncertainty from  calibration of 15\%. Column~15: Integrated spectral index calculated with the observed fluxes (columns 13 and 14).} 
\end{table*}
%%%%%%%

%%%%%%%%%%%%%%%%%%%%%
\section{Observations} \label{observations}
%%%%%%%%%%%%%%%%%%%%%

%%%%%%%%%%%
\begin{table*}
\label{table2}        
\centering                         
\begin{tabular}{c||c|c|c|c|c|c|c}   
\textbf{Name} & SpT& $\dot{M}$&$R_{\mathrm{cav}}$&$\lambda$ & $F$ (mJy) & $\alpha_{\mathrm{mm}}$& Reference\\
&&($M_\odot~$yr$^{-1}$)&(AU)&(mm)&$\pm\sigma$(mJy/beam)&\\
\hline
\hline
SR~21&G3&$<1 \times 10^{-9}$&36$^R$&0.88&$400\pm2.6$&$3.45\pm0.41$&(1)\\
&&&&3.30&$4.20\pm0.4$&&(2)\\
\hline
MWC~758&A8&$1 \times 10^{-8}$&73$^R$&0.88&$180\pm1.1$&$2.85\pm{0.58}$&(1)\\
&&&&2.70&$7.3\pm{1.4}$&&(3)\\
\hline
LkCa~15&K3&$2 \times 10^{-9}$&50$^R$&$0.88$&$410\pm0.9$&$2.82\pm{0.44}$&(1)\\
&&&&2.70&$17.4\pm{0.6}$&&(3)\\
\hline
GM~Aur&K5&$ 1\times 10^{-8}$&28$^R$&0.88&$640\pm3.5$&$2.94\pm{0.44}$&(1)\\
&&&&2.70&$23.7\pm{0.8}$&&(3)\\
\hline
DM~Tau&M1&$6 \times 10^{-9}$&19$^R$&0.88&$210\pm{1.3}$&$2.32\pm{0.44}$&(1)\\
&&&&2.70&$15.6\pm{0.4}$&&(3)\\
\hline
SR~24~S&K2&$1 \times 10^{-8}$&29$^R$&0.88&$550\pm{1.8}$&$2.29\pm{0.37}$&(1)\\
&&&&3.3&$26.6\pm{0.8}$&&(2)\\
\hline
DoAr~44&K3&$9 \times 10^{-9}$&30$^R$&0.88&$210\pm{2.7}$&$2.77\pm{0.38}$&(1)\\
&&&&3.3&$10.4\pm{0.5}$&&(2)\\
\hline
WSB~60&M4&$2 \times 10^{-9}$&15$^R$&0.88&$250\pm{1.3}$&$2.11\pm{0.37}$&(1)\\
&&&&3.3&$15.3\pm{0.5}$&&(2)\\
\hline
SZCha&K0G&$2.4 \times 10^{-9}$&18$^S$&1.2&$77.5\pm{20.3}$&$2.83\pm{0.87}$&(4)\\
&&&&3.0&$5.8\pm{0.5}$&&(4)\\
\hline
CSCha&K4&$1.2 \times 10^{-8}$&38$^S$&1.2&$128.4\pm{45.6}$&$2.85\pm{1.05}$&(4)\\
&&&&3.0&$9.4\pm{0.6}$&&(4)\\
\hline
CRCha&K0-K2&$8.8 \times 10^{-9}$&10$^S$&1.2&$124.9\pm{24.2}$&$3.28\pm{0.94}$&(4)\\
&&&&3.0&$6.2\pm{1.5}$&&(4)\\
\hline
J1604-2130&K2&$<1 \times 10^{-11}$&70$^R$&0.88&$164.0\pm{6.0}$&$3.20\pm{0.50}$&(5)\\
&&&&2.6&$5.1\pm{0.5}$&&(5)\\
\hline
RX J1615-3255&K5&$ 4 \times 10^{-10}$&30$^R$&0.88&$430.0\pm{2.9}$&$3.22\pm{0.41}$&(1)\\
&&&&3.2&$6.7\pm{0.6}$&&(6)\\
\hline
T~Cha&G2&$ 4 \times 10^{-9}$&15$^S$&1.2&$105.0\pm17.7$&$3.05\pm{0.79}$&(4)\\
&&&&3.0&$6.4\pm{1.0}$&&(4)\\
\hline
TW~Hydra&K7&$4 \times 10^{-10}$&4$^S$&0.87&$1340\pm130$&$2.61\pm{0.43}$&(7)\\
&&&&3.4&$41.0\pm{4.0}$&&(8)\\
\hline
HD163296&A1&$ 1 \times 10^{-11}$&$>25^R$&0.88&$1740\pm{120}$&$2.73\pm{0.44}$&(9)\\
&&&&2.8&$77.0\pm{2.2}$&&(10)\\
\hline
\end{tabular}
\caption{{\bf Additional data.} Column~1: Name of the target. Column~2: Spectral type. Column~3: Accretion rate. Column~4: Cavity radii ($R$ for resolved cavities and $S$ for cavities obtained from SED modelling). Column~5 and 6: Observed wavelength and the corresponding  millimetre fluxes. Column~7: Integrated spectral index calculated from the fluxes of column 6. Column~8: References: (1)\cite{andrews11}, (2)\cite{Ricci2010b}  (3)\cite{Guilloteau2011}, (4)\cite{ubach2012}, (5) \cite{mathews2012}, (6)\cite{lommen2010},  (7)\cite{andrews2012}, (8)\cite{wilner2003},  (9)\cite{qi2011} \& (10)\cite{isella2007}.}   
\end{table*}

\subsection{PdBI observations} \label{obs_pdbi}

The observations with PdBI were made on  20 and 21 September of 2012 under excellent weather conditions (seeing of $\sim$~1.7''), using only the WideX correlator. The sample consists of four transition disks: LkH$\alpha$~330,  UX~Tau~A,  LRLL~31,  and LRLL~67. These sources were selected based on the 880~$\mu$m flux information available \citep{andrews11b, espaillat12}, the lack of 3~mm data  in the literature, and their position in the sky. The 880~$\mu$m imaging of LkH$\alpha$~330 and UX~Tau~A has confirmed a dust-depleted cavity for these disks \citep{brown09, andrews11b}. There are no resolved images for LRLL~31 and LRLL~67. The available information about the position, stellar parameters, cavity radii, and millimetre fluxes of these targets are summarised in Table~\ref{table1}, with the corresponding references. 

The targets were observed in track-sharing mode with constant time intervals, inter-scanned  between the science target and quasars observations, including 0333+321 and 0507+179 for phase and amplitude calibration.  To calibrate the spectral response of the system,  3C84, 3C345, and 0234+285 were observed as bandpass calibrators. For the absolute flux calibration, MWC349 was used with a fixed flux value of 1.20~Jy  at 100.0~GHz. For the data reduction, the  Fourier inversion of the visibilities and the cleaning of the dirty image of the dust emission,  the Grenoble Image and Line Data Analysis Software (GILDAS) was used.  The observations were carried at a frequency of 100.0~GHz  with a bandwidth of $\sim~3.6$~GHz for continuum (3~mm) and with six antennas in the D-configuration. The D-antenna configuration is the most compact and  provides the lowest phase noise and highest sensitivity available using PdBI. In Appendix~\ref{appendix}, the 3mm maps for the observed targets, with the corresponding resolution and  noise level are presented. 

\subsection{Additional transition disks}

In addition to our own observations, we gathered available photometric data of other transition disks. The summary of this sample is in Table~\ref{table2} with the corresponding references. The target selection is based on transition disks for which the millimetre flux is known within a good uncertainty  at two different and well-separated (sub-)~millimetre wavelengths, preferably between $\sim880~\mu$m-1.2~mm and $\sim2.7-3.0$~mm. Shorter wavelengths are not taken to avoid possible contribution from optically thick dust. Longer wavelengths are not considered to minimise the contamination from ionised gas.

Most of our selected targets are disks around T Tauri stars, except for the Herbig disks MWC~758 and HD~163296. Table~\ref{table2} shows the spectral type, accretion rate, cavity radii, and  fluxes at millimetre wavelengths. The cavity radius reported in the table has either been resolved by continuum observations ($R$) or derived from SED ($S$) fitting. These values of the cavity radii from the SED fitting are degenerate and highly uncertain. Even more important, the cavity size measurements inferred by SED modelling and by mm-observations refer to two different kinds of grains, of micron and millimetre sizes respectively, which do not need to be spatially coexisting. An example where both measurements are available is the transition disk SAO 206462, for which the cavity size in micron size particles is smaller ($\sim$~28AU) than what is inferred for large dust grains from mm-observations  \citep[39 to 50 AU,][]{Garufi2013}.  Scattered light observations of  transition disks with high angular resolution are needed to confirm the cavity size and  the decoupling of large and small grains predicted indeed by dust evolution models to be observed \citep{juanovelar2013}.

\section{Results} \label{results}
In this section, we present the integrated spectral index for the transition disks sample and the correlation with the cavity size.

%%%%%%%%%%%%
%FIGURE 
%%%%%%%%%%%%
\begin{figure}
 \centering
   \includegraphics[width=9.0cm]{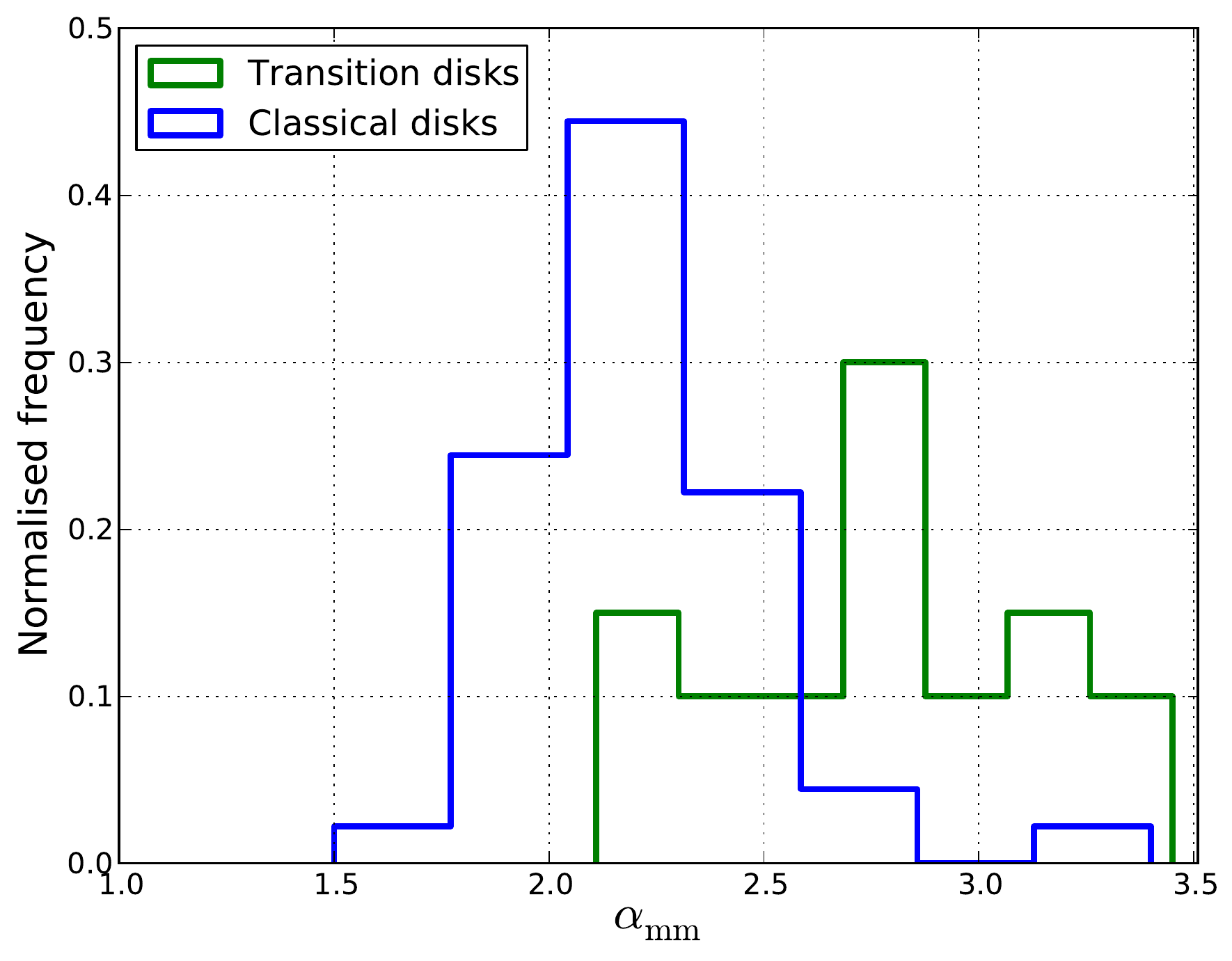}
   \caption{Comparison of the normalised distribution of the integrated spectral index for classical disks \citep[data from][]{Ricci2012} and transition disks (data in Tables~\ref{table1} and~\ref{table2}), using the same number of bins for each case.}
   \label{histogram}
\end{figure}

\subsection{Spectral Index}

At mm-wavelengths, the thermal emission  of a disk, integrated over its surface, is dominated by the emission of the outer, optically thin region. In the  Rayleigh-Jeans limit, the wavelength dependence of the integrated flux can be expressed as $F_\nu~\propto~\nu^{\beta_{\rm{mm}}+2}$, where $\beta_{\rm{mm}}$ is the opacity  index. It can be measured with multi-wavelength observations ($\beta_{\rm{mm}}~\equiv~d\ln\kappa/d\ln\nu$),  and the slope of the SED at long wavelengths can be approximated to  $\alpha_{\mathrm{mm}}~\approx~\beta_{\rm{mm}}+2$.  

Using the $880~\mu$m and 3~mm  fluxes of the observed targets and fluxes from literature, we calculated  the millimetre spectral indices $\alpha_{\mathrm{mm}}$. Tables~\ref{table1} and~\ref{table2} summarise our findings.  For calculating of the uncertainties of the spectral indices and the flux at 1~mm, besides the noise level or rms  ($\sigma$) of the observations, a systematic error from the flux calibration is included. It is taken to be 15\% of the total flux, since it is estimated from measurements with different instruments, such as CARMA, PdBI, and SMA  \citep[see e.g.][]{Ricci2011}.

%%%%%%%%%%%%
%FIGURE 
%%%%%%%%%%%%
\begin{figure*}
 \centering
   \includegraphics[width=16.0cm]{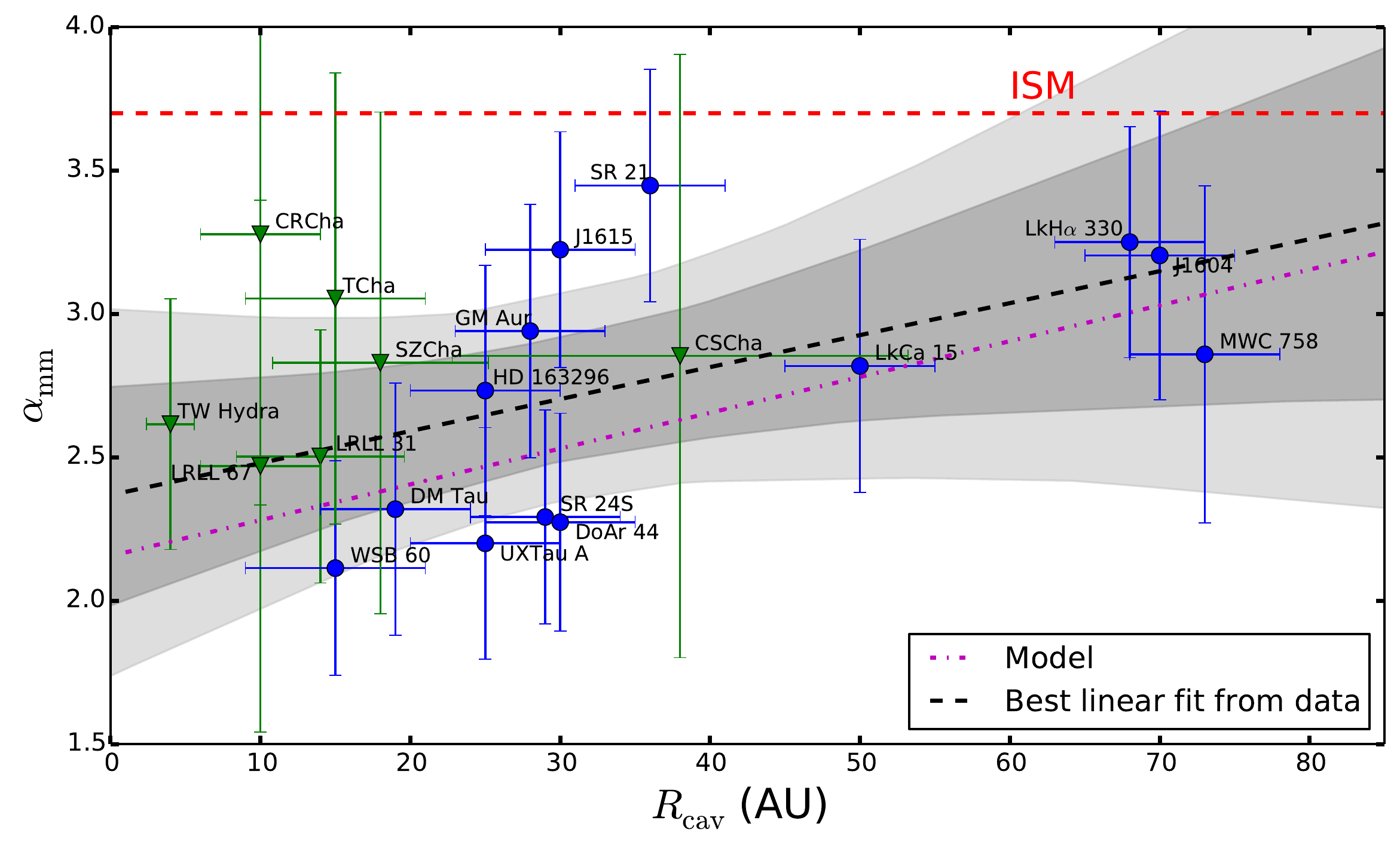}
   \caption{Spectral indices $\alpha_{\mathrm{mm}}$ vs the cavity radius for the observed transition disks. Data from our PdBI observations and available data from the literature  are plotted (data in Tables~\ref{table1} and \ref{table2}). Triangle points correspond to sources for which the cavity radius is inferred from SED modelling, while circle points correspond to resolved cavities by continuum observations. The uncertainties  on the  cavity radii are taken to be 40\% of the mean value in the case of SED modelling and $5~$AU for the resolved cavities. The dashed line corresponds to the linear regression from the data using the method introduced by \cite{kelly2007}. The shaded regions show the 95\% (2$\sigma$) confidence interval (lighter) and the 68\% (1$\sigma$) confidence region (darker)  on the regression line. The dot-dashed line represents the best linear regression from the models described in Sect~\ref{section_discussion}.}
   \label{alpha_flux_Rcav2}
\end{figure*}

Analysis of recent mm-observations at different wavelengths of several protoplanetary disks have shown that the opacity index may change with radius \citep{isella2010a, Guilloteau2011, perez2012, trotta2013}. For instance, \cite{Guilloteau2011} investigated a total of 23 disks in Taurus-Auriga region and found that for nine  sources,  $\beta_{\rm{mm}}$ increases with radius, implying that dust grains are larger in the inner disk ($\lesssim~100$~AU), where the gas density is higher, than beyond. If one assumes that transition disks had the same grain population before a clearing mechanism acted on them and also that the opacity index increases with radius \citep[e.g.][]{Guilloteau2011},  it is expected that  disks with  cavities lack large grains in the inner disk ($r~<~R_{\mathrm{cav}}$)  and  therefore have higher values of the integrated spectral index  than typical disks. 

To compare the integrated spectral index of regular protoplanetary  disks and transition disks, we use the data from  \citet{Ricci2012}, who presented millimetre fluxes and spectral indices of a total of $\sim$~50  disks in Taurus, Ophiuchus, and Orion star forming regions. The transition disk population includes disks in different star forming regions such as Taurus, Perseus, Lupus, Ophiuchus, and,  among others. Figure~\ref{histogram} shows the differences between the two distributions (plotted with the same number of bins), with a mean value of the spectral index $\bar{\alpha}_{\mathrm{mm}}^{\mathrm{PD}}=2.20\pm0.07$ for typical protoplanetary disks and  $\bar{\alpha}_{\mathrm{mm}}^{\mathrm{TD}}=2.70\pm0.13$ for transition disks. This comparison indicates that the millimetre index for transition disks is in average higher than for protoplanetary disks. 

In addition, to compare the two samples, we calculate the Kolmogorov-Smirnov (KS) two-sided test, finding that the maximum deviation between the cumulative distributions is $0.66$ with a significance level of $5.11 \times 10^{-6}$, i.e. with a very low probability ($\ll~1\%$) that the two samples are similar \footnote{For the histogram (Fig.~\ref{histogram}) and the K-S test, when the sources  are repeated  in both samples, we only consider them for the transition disk distribution. These targets are DM~Tau, GM~Aur, SR~21, and DoAr~44.} Moreover,  we  calculate the significance level 1000 times, considering random values for the uncertainties of the spectral index within the corresponding error bars for both samples, finding a very low mean probability that the two distributions are similar (less than $1\%$).

\subsection{Correlation of the spectral index and the cavity size}

The integrated spectral index  $\alpha_{\mathrm{mm}}$ of the four observed transition disks is lower than 3.5, although in some cases,  $\alpha_{\mathrm{mm}}$ is close to the mean value for the ISM. The large value of $\alpha_{\mathrm{mm}}$ for LkH$\alpha$~330 is  confirmed by \cite{isella2013}, who report CARMA observations at 1.3~mm of this transition disk and obtain  $\alpha_{0.88-1.3\mathrm{mm}}~=~3.6\pm0.5$. This can be  interpreted as no-grain growth, if one assumes constant values of $\beta_{\rm{mm}}$ over the whole disk.  From our observed targets, LkH$\alpha$~330 is the disk  with the  largest estimated inner hole, $\sim68$~AU \citep{andrews11}.  LRLL~31 and UXTau~A have lower spectral indices and smaller inner holes, 14 and 25~AU respectively \citep{andrews11, espaillat12}. The spectral index of LRLL67 has a large error,  preventing any conclusion on the relation between its integrated spectral index and the cavity radii.

Considering the sample of all transition disks (Tables~\ref{table1} and~\ref{table2}), Fig.~\ref{alpha_flux_Rcav2} shows the correlation between the spectral index and the cavity radius. For the  uncertainties  of the  cavity radius, we consider an uncertainty of 40\% for the cavity radii when they are derived from SED fitting.  This error may be underestimated due to the high degeneracy from the SED modelling.  When the cavities are resolved, an error of $\pm~5~$AU is considered \citep[$\sim$ 20\% of the resolution obtained from continuum observations at typical distances of $\sim$140~pc, see for example][]{andrews11}. Figure~\ref{alpha_flux_Rcav2} also shows the linear regression from the data using a Bayesian method introduced by \cite{kelly2007}, which includes measurement errors of the data in  both $R_\mathrm{cav}$ and $\alpha_\mathrm{mm}$, as well as intrinsic scatter. Fitting a linear relationship between the spectral index and the cavity radius, such that $\alpha_{\mathrm{mm}}=a\times R_{\mathrm{cav}}+b$, we find $a=0.011\pm{0.007}$ and $b=2.36\pm{0.28}$. We show the 95\% (2$\sigma$) confidence interval  and the 68\% (1$\sigma$) confidence region  on the regression line. All data from observations, when error bars are included, are contained within  95\% confidence interval. According to \cite{kelly2007}, 5000 iterations should be enough for the convergence of the linear coefficients. We assume 100000 iterations and took the last 95000 values to ensure  a good convergence of our estimations. This analysis shows that the probability of a positive correlation, considering uncertainties and scattering of the data,  is $\sim$95\%. We repeat this linear regression assuming a higher uncertainty (50\%) for the cavity radii obtained by SED modelling, finding that the probability of a positive correlation decreases by less than 1\%.

To determine whether the best fitting of the points in Fig.~\ref{alpha_flux_Rcav2} is linear, we use the $\chi^2$ test for several polynomial functions,  finding that the best fit to the data is obtained using linear fitting. The changes for $\chi^2$ for higher polynomial  orders are less than 1\% compared to the linear fitting. This suggests that any additional order above the linear polynomial, which we can take to model our observations, is over fitting the data.

The statistics remain low, but these observations suggest a positive dependence between $\alpha_{\mathrm{mm}}$ and the cavity radius. Observations with better resolution and sensitivity are needed  to refine this analysis and to determine an accurate relation between $R_{\mathrm{cav}}$ and $\alpha_{\mathrm{mm}}$.

\section{Discussion} \label{section_discussion}

Transition disks show a wide range of properties such as the extent of the observed inner holes from 4~AU to even more than 100~AU, relatively high accretion rates ($\sim~10^{-8}~M_\odot~\rm{yr}^{-1}$), and  averaged lifetimes ($\sim$~1-3~Myr), among others \cite[see][for a review] {williams11}. Recently,  different studies have combined various physical models to explain these properties. For instance, \cite{rosotti2013} combined X-ray photoevaporation models with planet-disk interactions to investigate the transition disks with accretion rates comparable to  classical T-Tauri stars and large inner holes. On the other hand, \cite{pinilla12a} studied how the dust evolves in a disk that  is perturbed by the presence of a massive planet. In that work, they demonstrate that the ring-shaped emission of some transition disks can be modelled using a single planet. The radial location of the mm-sized particles depends on the shape of the resulting gap, which is determined by the disk viscosity, mass, and location of the planet.  Other models such as a vortex generated by a dead zone \citep{Regaly2012}, a region in the disk where the viscosity drops down, may also explain some of the observed features of transition disks \citep{brown09, isella2013}. In this section, in order to  distinguish between the dust trapping scenario and a truncated disk case, we compared the observations presented in Sect.~\ref{results} with two different kind sof models. In the first set of models, we consider coagulation and fragmentation of dust particles in a disk with a cavity carved by a massive planet and with particle trap being at the outer edge of the cavity. In the second set of models, we assume  a truncated disk at different radii  without the computation of the dust evolution,  but a power law for the size distribution of particles.  

\subsection{Models of grain growth in transition disks}

%%%%%%%%%%%%
%FIGURE 
%%%%%%%%%%%%
\begin{figure}
 \centering
   \includegraphics[width=9.0cm]{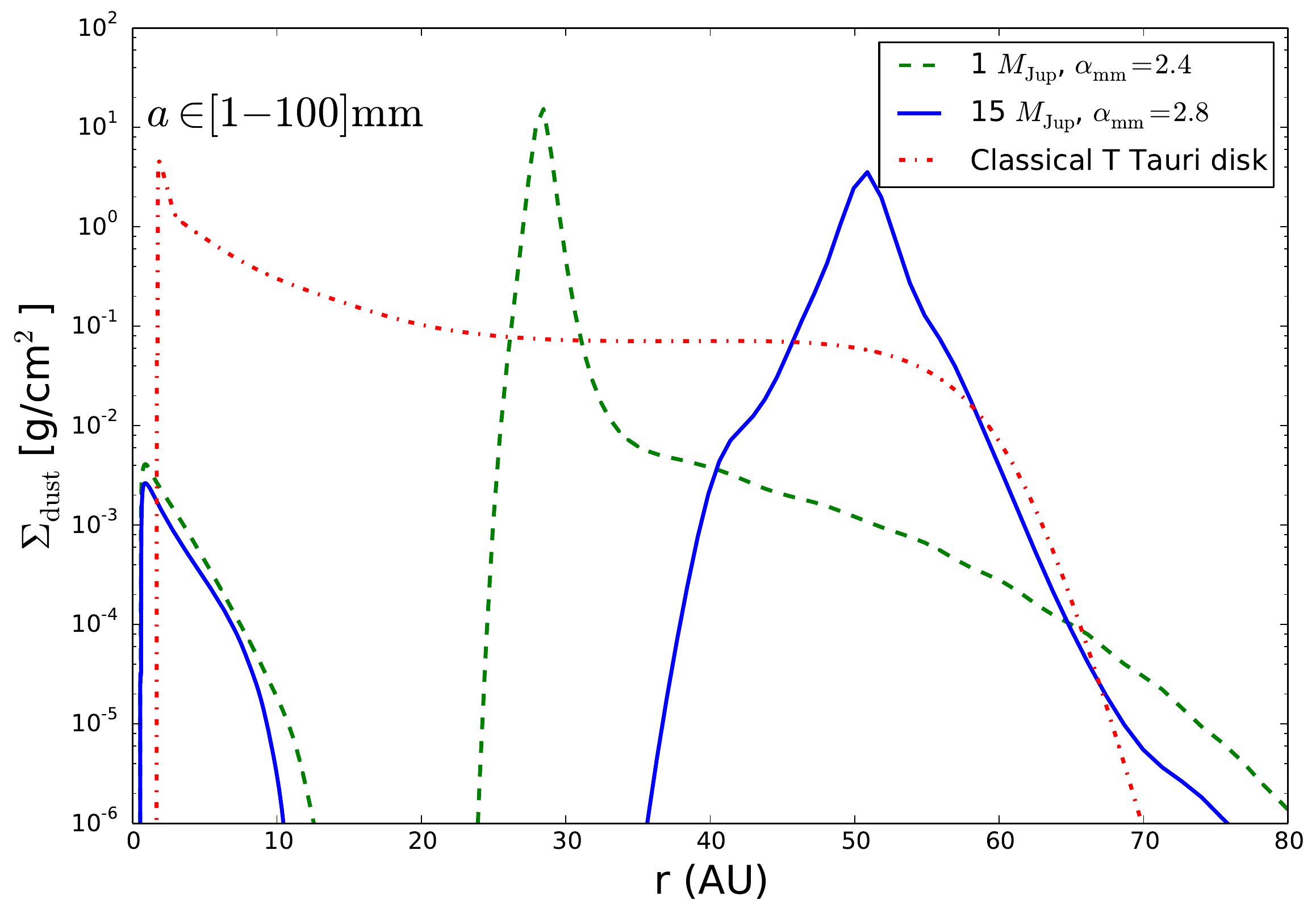}\\
   \includegraphics[width=9.0cm]{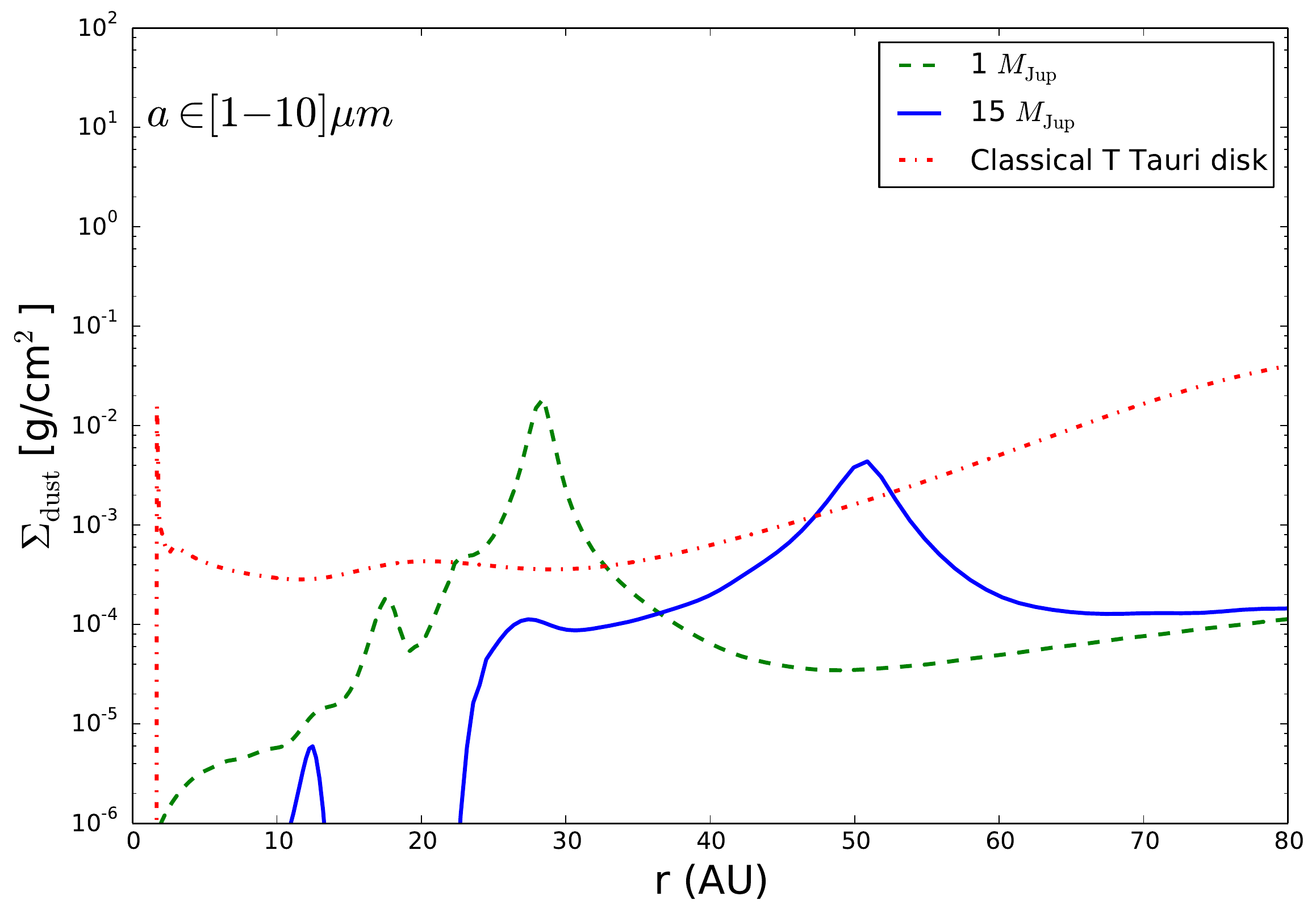}
   \caption{\emph{Top panel:} Dust surface density calculated from simulations in \cite{pinilla12a}  and considering  grain sizes ($a$) from 1 to 100~mm after 1~Myr of evolution. The disk hosts a massive planet at $20$~AU. In the case of 1~$M_{\mathrm{Jup}}$ (dashed line), the peak of dust surface density is at  $\sim~R_{\mathrm{cav}}~\sim~$30~AU, the spectral index is $\alpha_{1-3\mathrm{mm}}=2.4$, while for  15~$M_{\mathrm{Jup}}$ and the peak of the dust surface density at $\sim~R_{\mathrm{cav}}~\sim~$55~AU (solid line), the spectral index is $\alpha_{1-3\mathrm{mm}}=2.8$. The dash-dotted line represents the dust surface density with the same grain sizes in a protoplanetary disk without a planet, but after suppression of the radial drift. \emph{Bottom panel:} as top panel, but for micron-sized particles.}
   \label{dust_mm_grains}
\end{figure}

In our dust evolution models  \citep{birnstiel2010}, dust densities evolve due to radial drift, turbulent mixing, and  gas drag. At the same time, particles coagulate, fragment, and erode depending on their relative velocities. The sources for relative velocities are Brownian motion, turbulence, radial drift, and vertical settling. The final dust density distribution depends on disk parameters, such as the gas surface density, and viscosity;  particle parameters like fragmentation velocity and the intrinsic volume density of the spherical grains \citep[see][for more details]{birnstiel2010}.  In the case of a massive planet interacting with the disk, the gas surface density is calculated from simulations using the polar hydrodynamic code FARGO \citep{masset00}, after the disk evolves during $\sim~1000$ planet orbits. For the results present in this section, the disk parameters are assumed as in \cite{pinilla12a}, i.e., a disk around a Solar-type star, with a flared geometry and viscosity parameter $\alpha_{\mathrm{turb}}=10^{-3}$.  The disk mass is taken to be $0.05 M_{\odot}$ and the inner and outer radius of the disk change depending on the position of the planet, to have a disk extended from 1 to 150~AU.

When a single massive planet is interacting with the disk, it is expected that dust particles accumulate and grow to mm-sizes at distances that can be even twice or more than the star-planet separation. As a consequence, for a transition disk in which cavity is the result of a planet within the disk, the inner part ($r~<~R_{\mathrm{cav}}$) can be completely depleted of mm-grains depending on the planet mass. The top panel of Fig.~\ref{dust_mm_grains} shows the radial variation in the dust surface density $ \Sigma_{\mathrm{dust}}$ from the simulations presented in \cite{pinilla12a}, calculated only for millimetre particles i.e. 1-100~mm (top panel), for the cases of a 1 and 15~$M_{\mathrm{Jup}}$   planet located at 20~AU.  The case of a typical protoplanetary disk around a T Tauri star where the radial drift is reduced without any gas density perturbation (i.e. no pressure bump(s)) is also shown in Fig.~\ref{dust_mm_grains}.  The bottom panel of  Fig.~\ref{dust_mm_grains} shows for comparison the dust density for micron-sized particles (1-10~$\mu$m) under the same conditions. All  results are shown after 1~Myr of evolution.  The comparison between the two panels of Fig.~\ref{dust_mm_grains} shows that micron-sized particles are trapped less efficiently than mm-sized particles, leading to spatial  differences in the dust size distribution. 

The grain growth process under the presence of a pressure bump can lead  to significant radial variations of the opacity. When the inner disk is almost depleted of mm-grains, as we notice in Fig.~\ref{dust_mm_grains} for the 1 and 15$M_{\mathrm{Jup}}$ planet cases, the opacity index is high in the inner region and low in the outer part of the disk where the dust grains are trapped. This result contrasts with the findings on typical protoplanetary disks, where  $\beta_{\rm{mm}}$   increases with radius  \citep{Guilloteau2011, perez2012, trotta2013}. In fact, the radial variation of    $\beta_{\rm{mm}}$  for regular disks is expected to be inverse to the variation in the dust surface density of mm-grains, i.e. low values for $\beta_{\rm{mm}}$ in the inner region ($r\lesssim 70$~AU) and higher in the outer part. 

\subsection{Comparison with observations} \label{comparison_obs}

To calculate the disk-integrated spectral index $\alpha_{\mathrm{mm}}$ for a transition disk in this model, we considered dust density distributions after 1~Myr of dust evolution as in Fig.~\ref{dust_mm_grains}. We calculated the opacities $\kappa_\nu$ for each grain size at a given frequency $\nu$, assuming the optical constants for magnesium-iron silicates \citep{jaeger1994, dorschner1995} from the Jena database\footnotemark{}\footnotetext{$\textrm{http://www.astro.uni-jena.de/Laboratory/Database/databases.html}$} and using Mie theory.  The grid for the particle sizes runs from 1~$\mu$m to 200~cm. Once the opacities were calculated, the optical depth $\tau_\nu$ was computed as

\begin{equation}
	\tau_\nu=\frac{\sigma(r,a) \kappa_\nu}{\cos i},
  \label{eq:opt_depth}
\end{equation}

\noindent where $\sigma(r,a)$ is the total dust surface density and $i$ the disk inclination. The flux of the disk at a given frequency $F_\nu$ is therefore

\begin{equation}
	F_\nu=\frac{2\pi\cos{i}}{d^2}\int_{R_{\mathrm{in}}}^{R_{\mathrm{out}}} B_\nu(T(r)) [1-e^{-\tau_\nu}] r dr,
  \label{eq:flux}
\end{equation}

\noindent with $d$ the distance to the source and $B_\nu(T(r))$ the Planck function for a given temperature profile $T(r)$. The temperature is calculated self-consistently between the hydrodynamical simulations, the dust evolution, and the calculation of the millimetre fluxes. The millimetre emission computed from the dust density distribution that results from the dust evolution models is optically thin. Thus, the variations of the integrated millimetre spectral index exclusively come from different dust size distributions in each case, since in our models the disk is in the Rayleigh-Jeans regime for all radii.

All spectral indices are calculated assuming the same disk and dust parameters. These parameters are a disk around a typical T Tauri star of $1$~$M_{\odot}$ mass, whose mass is $M_{\mathrm{disk}}=0.05 M_{\odot}$, a viscosity parameter of $\alpha_{\mathrm{turb}}=10^{-3}$, fragmentation velocities of particles of 10~m~s$^{-1}$, and the same initial dust surface density. The intrinsic volume density $\rho_s$ of a spherical grain is taken to be 1.2~g~cm$^{-3}$. The effect of each of these parameters on the opacity index was investigated in detail by  \cite{Birnstiel2010b} and \cite{pinilla2013}, and it will be discussed later. We consider $d=140$~pc and $i=0$ (face-on disks).  The cavity radius depends on the mass and planet location, and we focus on these two variables to study the effect that the cavity shape has on the value of $\alpha_{\mathrm{mm}}$.

\paragraph{\emph{Effect of the planet mass\\} }

For our research, we calculated the millimetre fluxes at 1 and 3~mm using Eq~\ref{eq:flux}. In this case, we considered two cases of 1~$M_{\mathrm{Jup}}$  and  15~$M_{\mathrm{Jup}}$ located at the same orbital distance (20AU as in Fig.~\ref{dust_mm_grains}). The disk integrated spectral index for the  1~$M_{\mathrm{Jup}}$ and $R_{\mathrm{cav}}~\sim~$30~AU is $\alpha_{1-3\mathrm{mm}}=2.4$, while for 15~$M_{\mathrm{Jup}}$ and $R_{\mathrm{cav}}~\sim~$55~AU, it is $\alpha_{1-3\mathrm{mm}}=2.8$. The cavity radius  $R_{\mathrm{cav}}$ is taken to be where the peak of the dust surface density is located and observed at mm-wavelenghts. Increasing the cavity radii by more than 20~AU using a more massive planet increases the value of the integrated spectral index as well. 

When a single massive planet interacts with the disk, the size of the  resulting cavity at mm-wavelengths can be written in terms of the Hill planet radius $r_H$, defined as $r_H~=~r_p~(M_p/3M_\star)^{1/3}$, where $r_p$ and $M_p$ are the radial position and mass of the planet. For $1~M_{\mathrm{Jup}}<Mp<5~M_{\mathrm{Jup}}$, the mm-emission is expected to be at  $\sim7~r_H$, while for a disk hosting a more massive planet, the emission is expected at  $\sim10~r_H$ \citep{pinilla12a}. Increasing the planet mass  is  therefore not as effective as increasing the planet location to create wide cavities. For this reason a factor of 15, for example, in the planet mass has a moderate effect on the integrated spectral index.

\paragraph{\emph{Effect of the planet location\\} }
To investigate the influence of the planet location, we restrict our analysis to the the case of  1~$M_{\mathrm{Jup}}$ located at six different positions  from $r_p=10$~AU to  $r_p=60$~AU. In this case,  the mm-cavity is in a range of $R_{\mathrm{cav}}=20-90$~AU. Figure~\ref{alpha_nogrowth} shows the integrated spectral index vs. the cavity radius from simulations. The results confirm that indeed $\alpha_{\mathrm{mm}}$ linearly increases with the cavity radius.  Figure~\ref{alpha_nogrowth}  also shows the best linear fit of the model prediction, which minimises the squared error $\chi^2$: $\alpha_{\mathrm{1-3mm}}=0.012\times R_{\mathrm{cav}}+2.15$. This fit is over-plotted in Fig.~\ref{alpha_flux_Rcav2} and shows that this model fit explains the observations well.

The increase in  $\alpha_{\mathrm{mm}}$ with $R_{\mathrm{cav}}$ is because the disk integrated emission is dominated by the dust at $R_{\mathrm{cav}}$ in the pressure trap, in which the maximum size of particles decreases when   $R_{\mathrm{cav}}$ is further from the star. When the radial drift is reduced in a particle trap, turbulent motion is the main source of destructive collisions. As a result, the maximum size of the particles $a_{\mathrm{max}}$ is reached when the fragmentation velocity of the particles $v_f$ is equal to the turbulent relative velocity. In this case, $a_{\mathrm{max}}$ scales with the gas surface density $\Sigma_g$ as  \citep{birnstiel2010}

\begin{equation}
a_{\mathrm{max}}\propto \frac{\Sigma_g}{\alpha_{\mathrm{turb}}\rho_s}\frac{v_f^2}{c_s^2},
\label{a_max}
\end{equation}

\noindent where $\alpha_{\mathrm{turb}}$ is a turbulence parameter, $\rho_s$ the mass density of a single grain, and $c_s$ the sound speed.  Equation~\ref{a_max} implies that the maximum size  that particles reach in the coagulation/fragmentation cycle decreases with radius as the gas surface density does. Moreover, the size of the particles that are perfectly trapped ($a_{\mathrm{critical}}$) in a pressure bump ($P$) also scales with the gas surface density as \citep{pinilla2012}

\begin{equation}
a_{\mathrm{critical}}\propto\frac{\alpha_{\mathrm{turb}}\Sigma_g}{\rho_s |d\log P/ d \log r |}.
\label{a_max1}
\end{equation}

Therefore, for disks with similar parameters, but with the pressure bump located further out from the star (wider cavity), $a_{\mathrm{max}}$ and $a_{\mathrm{critical}}$ are expected to be smaller. For instance, for the parameters considered in this model, we have  $a_{\mathrm{max}}\sim~10$~cm  at 50~AU, while at 100~AU $a_{\mathrm{max}}\sim~1$~cm, leading to higher values of the integrated spectral index as shown in  Fig.~\ref{alpha_nogrowth}. 

\paragraph{\emph{Effect of other parameters\\} }
Disk and stellar parameters also influence the integrated spectral index, such as disk and stellar mass, turbulence, and disk size.  For instance, if the disk mass decreases, this also decreases the millimetre fluxes and the maximum size of particles (because $\Sigma_g$ also decreases,  if the gas surface density profile remains the same, Eq.~\ref{a_max}). As a consequence, the integrated spectral index is expected to be higher for low mass disks \citep{Birnstiel2010b}. Variations in the gas surface density profile, keeping the same disk mass, do not have a significant influence on $\alpha_{\mathrm{mm}}$ \citep{pinilla2013}. On the other hand, radial drift is more efficient for particles around low-mass stars, and millimetre grains around low-mass stars stay in the outer regions during shorter times than around solar-type stars \citep{pinilla2013}.  Disk turbulence also affects the maximum grain size, and low values of  $\alpha_{\mathrm{turb}}$ allow  larger grains and decrease the spectral index. Similarly, if the threshold velocity for destructive collision increases i.e. if $v_f$ is higher, then particles would reach larger sizes  (Eq.~\ref{a_max}). As a consequence, varying those parameters would change the slope and the intercept of the linear relation and broaden the correlation found by dust evolution modelling.

\subsection{Truncated disk}

To compare our previous results that include the computation of the dust coagulation/fragmentation with a truncated dust-rich disk, we consider a power-law grain size distribution given by $n(a)\propto a^{-s}$ with $a_{\mathrm{min}}=1\mu$m, $a_{\mathrm{max}}$ scaling as in Eq.~\ref{a_max}, and $s$ taken to be $s=[3, 3.5, 4]$. For  $T$, $\alpha_{\mathrm{turb}}$, $v_f$, and $\rho_s$, the same values are taken as in the dust evolution models. The gas surface density is taken as in \cite{Lyden-Bell1974},

\begin{equation}
	\Sigma_g(r)=\Sigma_0 \left(\frac{r}{r_c}\right)^{-\gamma}  \exp \left[-\left(\frac{r}{r_c}\right)^{2-\gamma}\right],
	\label{density}
\end{equation}

%%%%%%%%%%%%
%FIGURE 
%%%%%%%%%%%%
\begin{figure}
 \centering
   \includegraphics[width=9.0cm]{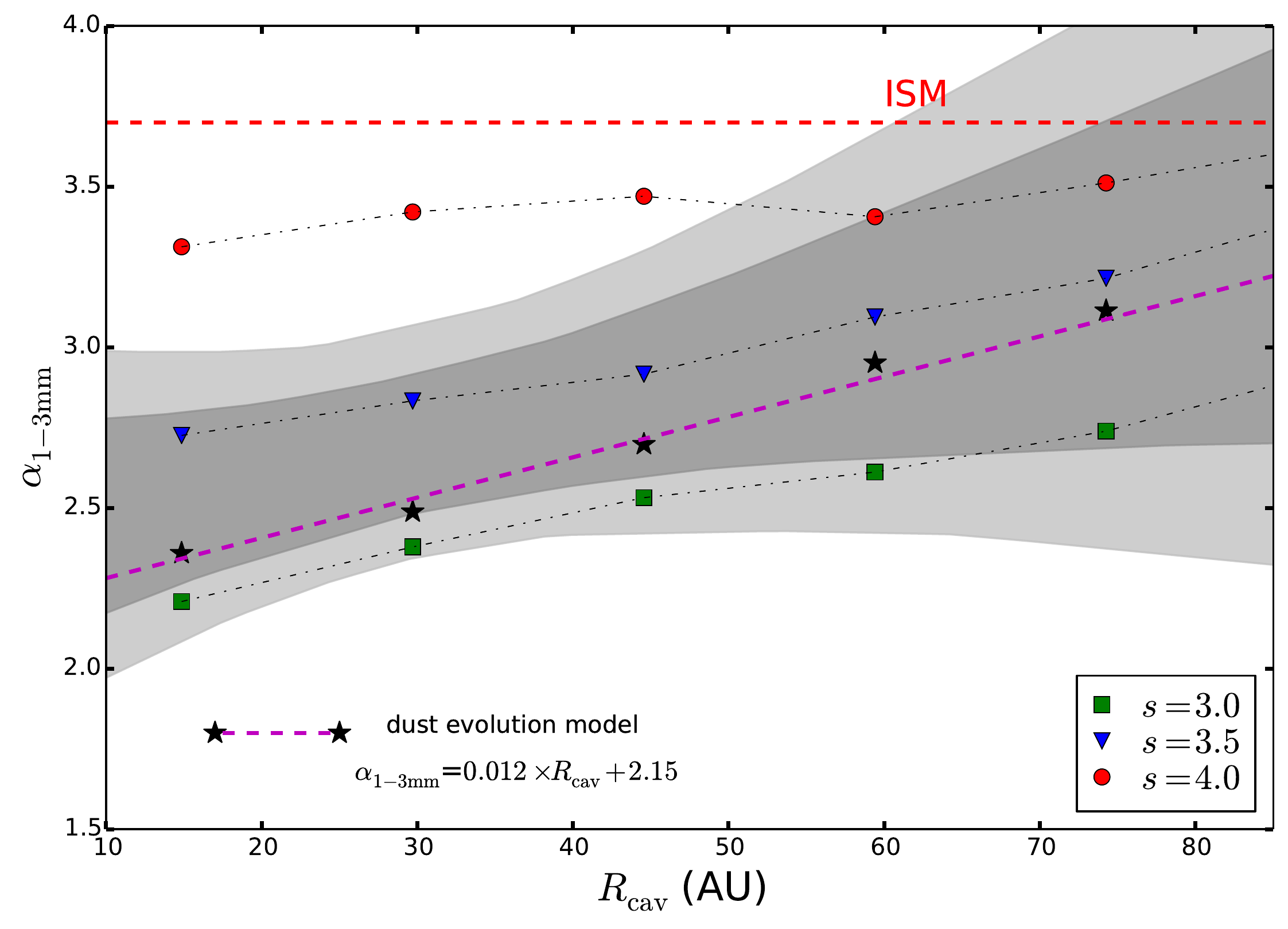}
   \caption{Disk-integrated spectral index $\alpha_{1-3\mathrm{mm}}$ vs the cavity radius for two different kind of models: (1) dust evolution in a disk holding a 1~$M_{\mathrm{Jup}}$ located at different distances and created different cavity size (stars in the plot). For this case, the best linear fit to this model is also plotted (dashed line). (2) Models without considering dust evolution, but a dust-rich disk, assuming a power-law grain size distribution ($n(a)\propto a^{-s}$), with $a_{\mathrm{min}}=1\mu$m, $a_{\mathrm{max}}$ given by Eq.~\ref{a_max},  three different values for $s$ ($s=3.0$, square points; $s=3.5$ triangle points, and $s=4.0$ circle points), and different truncated radii.  The 1 and 2~$\sigma$ regions from observations (Fig.~\ref{alpha_flux_Rcav2}) are also plotted.} 
   \label{alpha_nogrowth}
\end{figure}

\noindent where $r_c$ is the characteristic radius assumed to be $60$~AU, $\Sigma_0$ is calculated such that the disk mass is $0.05~M_{\odot}$, and $\gamma$ is equal to $1$.  Since the shape of the gas surface density does not significantly influence the integrated spectral index, variations in the characteristic radius and $\gamma$ are not considered.  To mimic different cavity sizes and study the effect of just removing the optically thick inner disk on the integrated spectral index, an artificial cut at different radii is done, inside which the disk is empty of dust.  Using the same procedure of Sect.~\ref{comparison_obs} to calculate the millimetre fluxes, Fig.~\ref{alpha_nogrowth} shows the  correlation between the integrated spectral index and disks with different cavity sizes. 

Increasing the slope of the size grain distribution $s$, implies less large grains and higher values of  $\alpha_{\mathrm{mm}}$. For this reason, in the cases of $s=3.5$ and $s=4$, the spectral index is high compare with the dust evolution models.  Only in the case of $s=3.0$, does $\alpha_{1-3\mathrm{mm}}$ reach low values and converge almost to the same value  as in the dust evolution models for a disk without a cavity ($R_\mathrm{cav}\rightarrow0$,  see Fig.~\ref{alpha_nogrowth}).  Considering different values for $M_{\mathrm{disk}}$, $\alpha_{\mathrm{turb}}$, and $v_f$, the corresponding variations in the spectral index are as expected from the models that include the computation of dust evolution. This is because the maximum grain size is computed following Eq.~\ref{a_max}, which assumes that turbulence causes destructive collisions, as in the case of the coagulation/fragmentation calculation with reduced radial drift.

The two frameworks presented in this work, the planet-disk interaction with dust evolution and a truncated disk with a power law size distribution,  remain  possible explanations for the positive trend obtained by observations (Fig~\ref{alpha_nogrowth}). However,  combinations of  scatter light and millimetre wavelength observations, with high angular resolution,  of transition disks will distinguish between the two scenarios \citep{juanovelar2013}.

\section{Summary and conclusions} \label{conclusion}

We have reported 3mm observations carried out with PdBI of four transition disks LkH$\alpha$~330, UX-Tau~A, LRLL~31, and LRLL~67.  Using previous observations of these targets at 880~$\mu$m, we calculated the spectral index, and found that for  these sources the values for $\alpha_{\mathrm{mm}}$ are lower than typical values of the ISM dust. However, the disk-integrated spectral index $\alpha_{\mathrm{mm}}$ is close to the value for ISM-dust ($\sim$~3.7) for LkH$\alpha$~330 ($\alpha_{\mathrm{mm}}=3.25\pm{0.40}$), which  is also the target with the largest inner hole \citep[$\sim$~68~AU,][]{andrews11b}. The sources UX-Tau~A and LRLL~31 have smaller cavity radii of 25~AU and 14~AU, respectively, and lower integrated spectral indices, $\alpha_{\mathrm{mm}}=2.20\pm{0.40}$ for UX-Tau~A  and $\alpha_{\mathrm{mm}}=2.50\pm{0.44}$ for LRLL~31. Owing to the large uncertainties on the millimetre flux of  LRLL~67, there is no conclusive correlation between the integrated spectral index and its cavity radius.

Recent observations at different wavelengths have shown that the opacity may  change radially throughout the disk \citep{isella2010a, Guilloteau2011, perez2012, trotta2013}. Studies that combine dust evolution and planet-disk interaction for transition disks \citep{pinilla12a} indicate  that the dust size distribution may present strong spatial variations. In this picture, the inner part is depleted of mm grains  leading to radial variations of the dust opacity. Thus,  the opacity index has higher values in the inner regions than  the outer regions, in contrast to observations of classical protoplanetary disks. From  models that combine hydrodynamical simulations and dust evolution models for the planet-related cavities, a linear relation between the disk integrated spectral  index and the cavity radius is inferred, because the millimetre emission is dominated by the dust at the outer edge of the cavity, i.e. in the pressure bump. This suggests that the spectral index is higher for larger cavities. 

Since our own sample is too small to draw statistically significant conclusions, we compiled all the known millimetre fluxes from transition disk sources for which these data exist at least for two separated millimetre wavelengths. Together with our observations, we collecteded a total of twenty transition disks for which the cavity radii are known either by SED modelling or spatially resolved observations,  and we calculate the disk integrated spectral indices. We found that  the integrated spectral index is higher for transition disks than for regular protoplanetary disks. 

Moreover, the probability of a positive trend between the spectral index and cavity size is $\sim$95\% (Fig.~\ref{alpha_flux_Rcav2}). Two models are considered in this work to explain this positive correlation: first of all, coagulation and fragmentation of dust in a disk whose cavity is formed by a massive planet located at different positions; second, a disk with different truncation radii and with a power-law dust-size distribution where turbulence sets the maximum grain size. Both models can explain the trend, and multi-wavelength observations with high angular resolution  are needed to distinguish between the two cases.  

Observations with new telescope arrays such as ALMA, EVLA, and NOEMA will provide  enough sensitivity and angular resolution to detect the small cavities of many transition disks and constrain  the dust opacity distribution of those disks in more detail.  Together with scattered imaging, these observations will help us understand the correlation between the spectral index and disk properties, such as the cavity size, hence, the environment where the planets are born.

\begin{acknowledgements}
The authors are very thankful to J.M. Winters for  his help understanding GILDAS for the PdBI data reduction. We acknowledge S. Andrews, N.~van der Marel, and E. van Dishoeck for the helpful comments and feedback on the manuscript. TB acknowledges support from NASA Origins of Solar Systems grant NNX12AJ04G. This publication is based on observations carried out with the IRAM Plateau de Bure Interferometer. IRAM is supported by INSU/CNRS (France), MPG (Germany), and IGN (Spain).

\end{acknowledgements}

\appendix
\section{Observed targets} \label{appendix}

\paragraph{\emph{LkH$\alpha$ 330:}} a G3 young star located in the Perseus molecular cloud at a distance of $\sim$250~pc. The first SMA observations were done by \cite{brown2007, Brown2008} and then by \cite{andrews11}. From these observations,  a gap of $\sim$68~AU radius at the continuum was resolved. Assuming azimuthal symmetry, they derived an inclination between 42-84$^\circ$. Recently, this disk was observed at 1.3~mm by \cite{isella2013} with the CARMA interferometer. They achieved an angular resolution of 0.35'', resolving a lopsided ring at $\sim$~100~AU from the central star and with an intensity azimuthal variation of $\sim$~2 and assuming a disk inclination of 35$^\circ$. 

\paragraph{\emph{UX~Tau~A:}}   a G8 primary star of a multiple system located in the region of Taurus at a distance of $\sim$~140~pc. \cite{andrews11b} report the first resolved observations of this disk with SMA, with a resolution of 0.31'' x 0.28''.  A gap of 25~AU of radius was found.  A disk inclination of 52$^\circ$ was inferred from SED fitting, although this value is highly degenerated. UX~ Tau is a triple system, UX~Tau B and C are classified as weak-lined T~Tauri stars, and A is the only classical T Tauri star, which is seven times more brighter than component B and 24 times brighter than C \citep{furlan2006}. The component A is separated by 5.86'' from B and 2.63'' from C. However, UX~Tau~A is the only component of the system with  evidence of a circumstellar disk \citep{white2001, mcabe2006}.

\paragraph{\emph{LRLL~31:}} a G6 young star located in the IC~348 star forming region at a distance of $\sim$~315~pc. The only millimetre flux densities reported in the literature are by \cite{espaillat12} using the compact configuration (C) of the SMA (with a maximum beam size of $\sim$~2.12'' x 1.90''). From the SED fitting, they classified this disk as a  ``pre-transition disk'', with an optically thick inner disk of a radius of 0.32~AU, separated by an optically thin gap of a radius of 14~AU.

\paragraph{\emph{LRLL~67:}} a M0.75 young star, which is also located in the IC~348 region. It was observed with SMA by \cite{espaillat12} with the same antenna configuration as LRLL~31.  From the SED modelling, this disk was classified by  as a transition disk with a cavity of 10~AU of radius.

\begin{figure*}
  \centering
  \tabcolsep=0.10cm 
   \begin{tabular}{cc}
   \includegraphics[width=8.5cm]{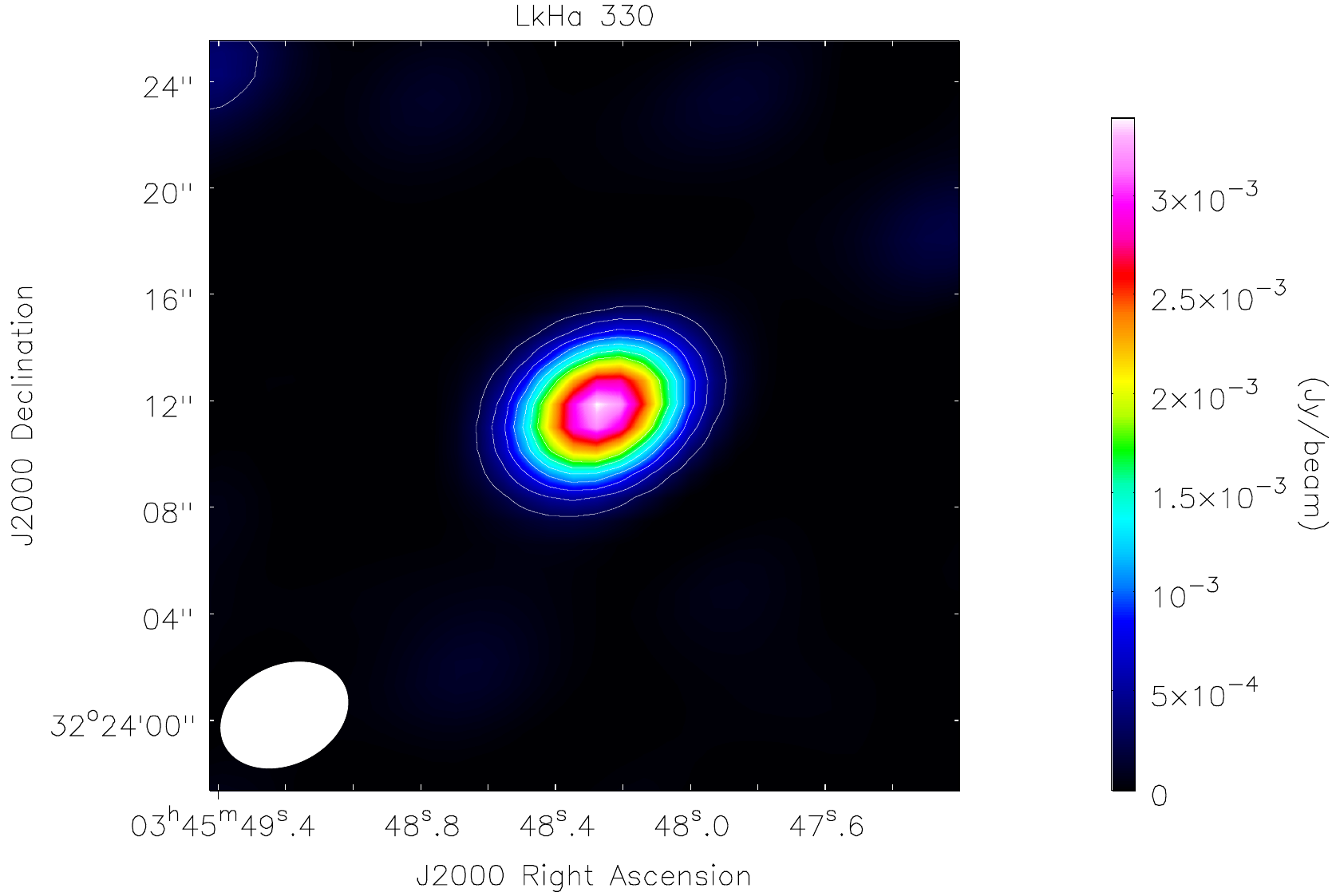} & 
   \includegraphics[width=8.5cm]{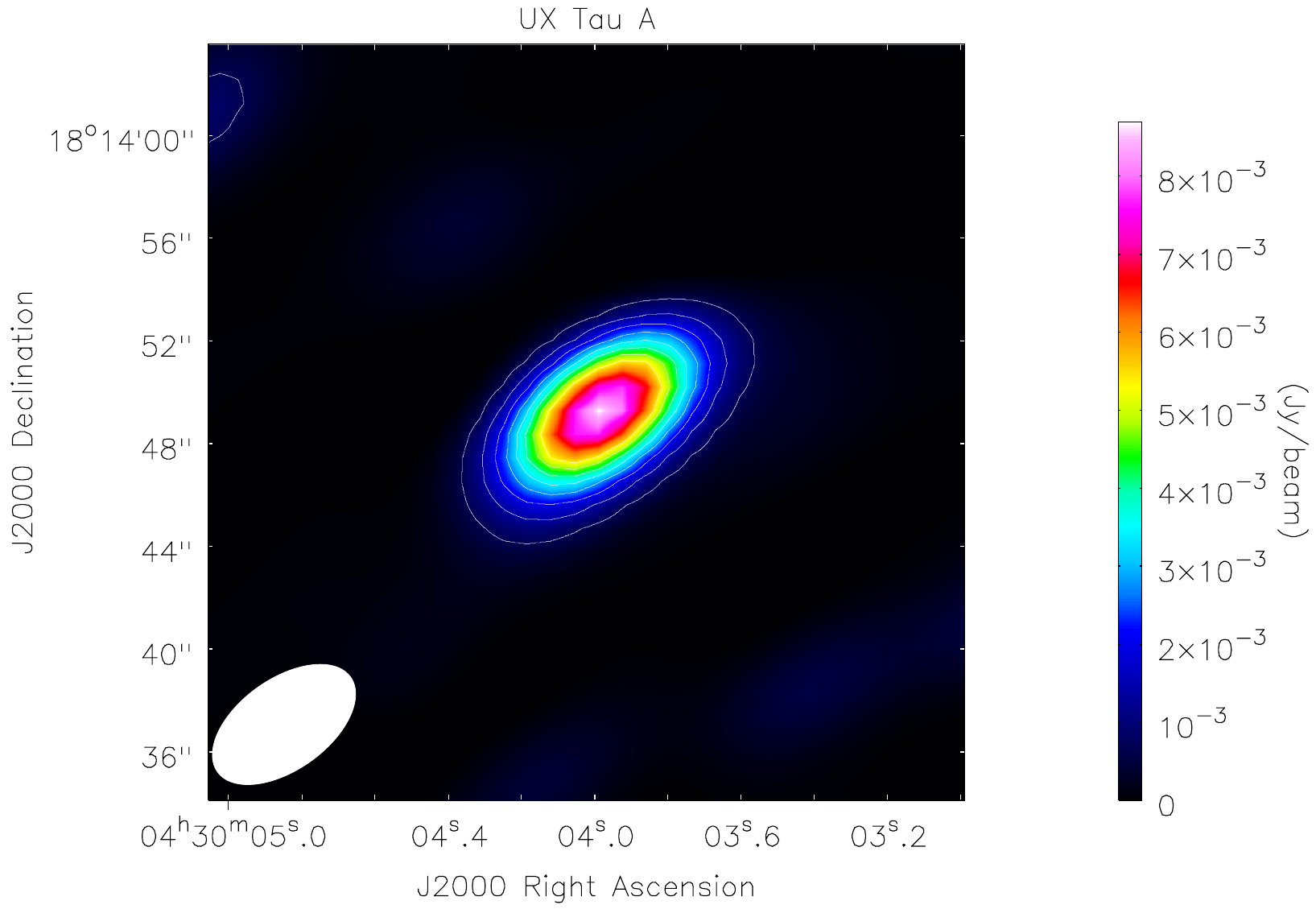}\\
   \includegraphics[width=8.5cm]{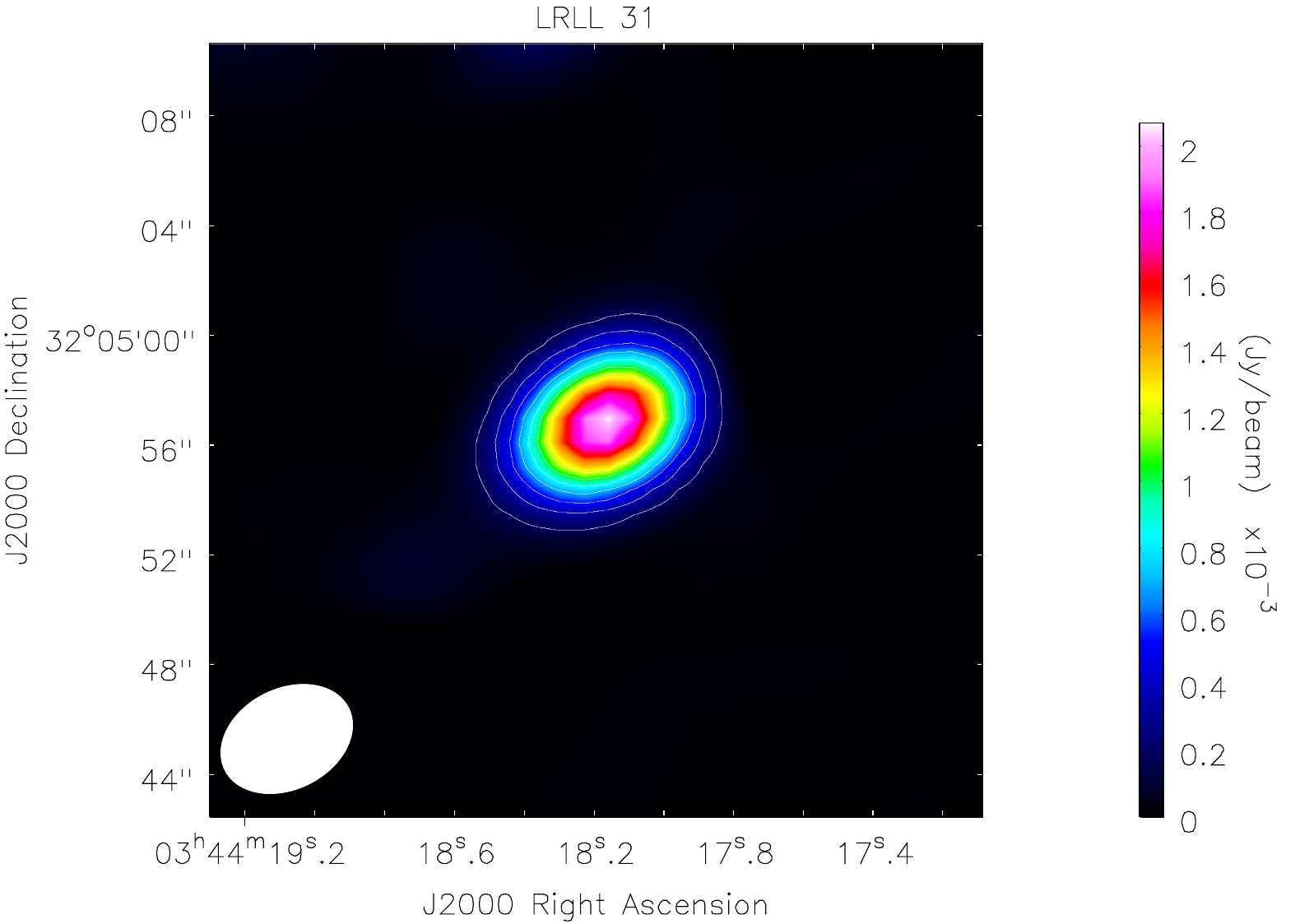} & 
   \includegraphics[width=8.5cm]{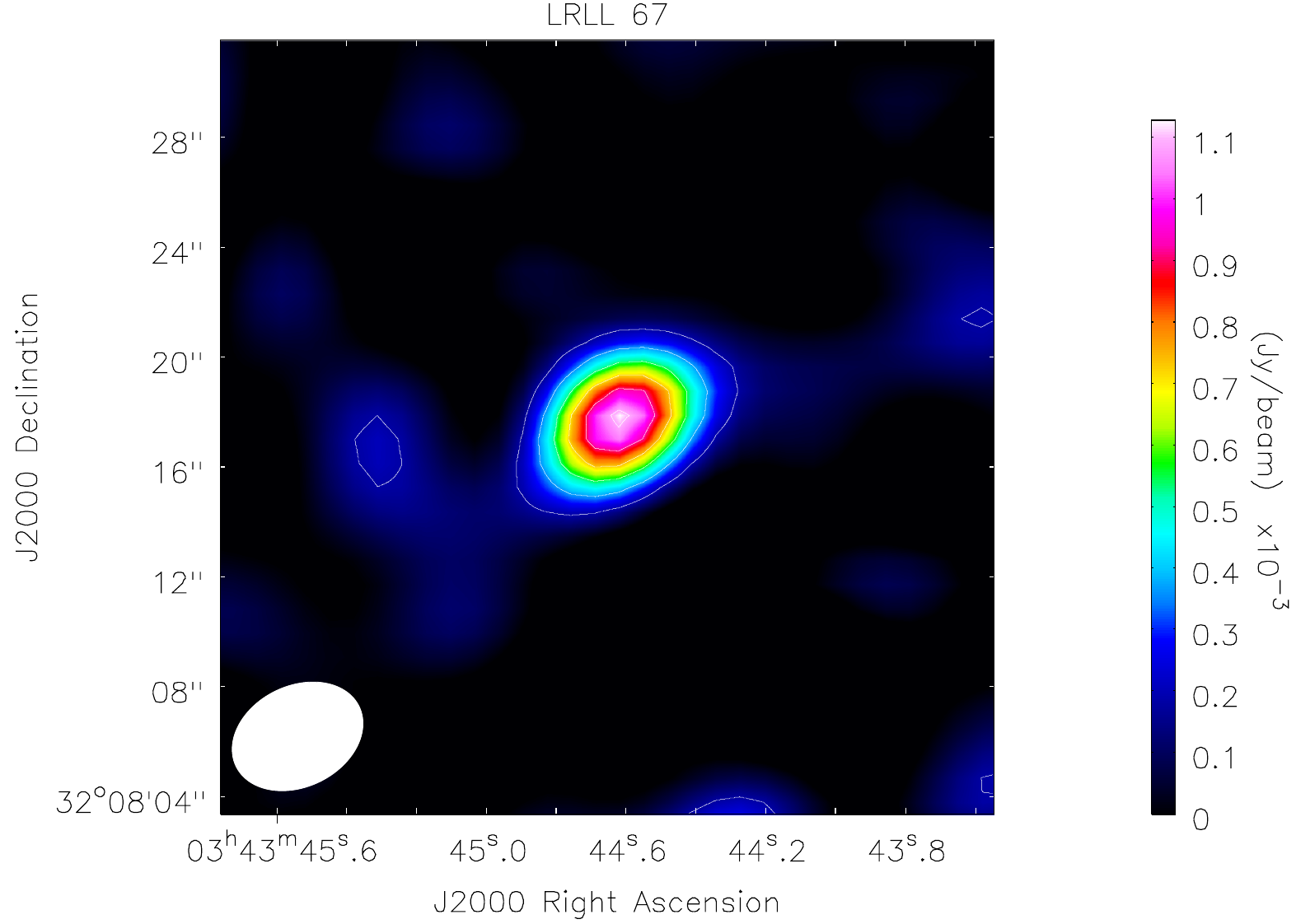}\\
   \end{tabular}	
\caption{Continuum maps  at 100.50~GHz (3~mm) obtained with the PdBI observations with contour lines every $2\sigma$. \emph{Upper left panel:} LkH$\alpha$~330 with $\sigma=0.12$mJy/beam, beam size of 5.01''$\times$3.69'' and position angle of the major axis of 117$^\circ$. \emph{Upper right panel:} UX~Tau~A  with $\sigma=0.33$mJy/beam, beam size of 6.36''$\times$3.54'' and position angle of the major axis of 125$^\circ$. \emph{Bottom left panel:} LRLL~31 with $\sigma=0.08$mJy/beam, beam size of 5.05''$\times$3.66'' and position angle of the major axis of 117$^\circ$.  \emph{Bottom left panel:} LRLL~67 with $\sigma=0.09$mJy/beam, beam size of 4.98''$\times$3.67'' and position angle of the major axis of 116$^\circ$.  The beam is shown in the  bottom right of each image.}
 \label{Fig:maps}
\end{figure*}

Figure~\ref{Fig:maps} shows the 3~mm continuum maps obtained with the PdBI observations for the four targets. The contour lines are every 2$\times\sigma$ with the corresponding $\sigma$  for each source. The values of the noise level ($\sigma$) are given in Table~\ref{table1} for each target. 

%%%%%%%%%%%%
%FIGURE
%%%%%%%%%%%%

\bibliographystyle{aa}

\bibliography{pdbi_pinilla_two_columns.bbl}

\begin{thebibliography}{80}
\expandafter\ifx\csname natexlab\endcsname\relax\def\natexlab#1{#1}\fi

\bibitem[{{Andrews} {et~al.}(2011{\natexlab{a}}){Andrews}, {Rosenfeld},
  {Wilner}, \& {Bremer}}]{andrews11}
{Andrews}, S.~M., {Rosenfeld}, K.~A., {Wilner}, D.~J., \& {Bremer}, M.
  2011{\natexlab{a}}, \apj, 742, L5

\bibitem[{{Andrews} {et~al.}(2011{\natexlab{b}}){Andrews}, {Wilner},
  {Espaillat}, {Hughes}, {Dullemond}, {McClure}, {Qi}, \& {Brown}}]{andrews11b}
{Andrews}, S.~M., {Wilner}, D.~J., {Espaillat}, C., {et~al.}
  2011{\natexlab{b}}, \apj, 732, 42

\bibitem[Andrews et al.(2012)]{andrews2012} Andrews, S.~M., Wilner, 
D.~J., Hughes, A.~M., et al.\ 2012, \apj, 744, 162

\bibitem[Ataiee et 
al.(2013)]{ataiee2013} Ataiee, S., Pinilla, P., Zsom, A., et al.\ 2013, \aap, 553, L3


\bibitem[Biller et al.(2012)]{biller2012} Biller, B., Lacour, S., 
Juh{\'a}sz, A., et al.\ 2012, \apjl, 753, L38

\bibitem[Birnstiel et 
al.(2010)]{birnstiel2010} Birnstiel, T., Dullemond, C.~P., \& Brauer, F.\ 2010, \aap, 513, A79

\bibitem[Birnstiel et 
al.(2010b)]{Birnstiel2010b} Birnstiel, T., et al.\ 2010b, A\&A, 516,
L14

\bibitem[Birnstiel et 
al.(2013)]{birnstiel2013} Birnstiel, T., Dullemond, C.~P., \& Pinilla, P.\ 2013, \aap, 550, L8


\bibitem[Brauer et 
al.(2007)]{brauer07} Brauer, F., Dullemond, C.~P., Johansen, A., et al.\ 2007, \aap, 469, 1169

\bibitem[Brown et al.(2007)]{brown2007} Brown, J.~M., Blake, 
G.~A., Dullemond, C.~P., et al.\ 2007, \apjl, 664, L107

\bibitem[Brown et al.(2008)]{Brown2008} Brown, J.~M., Blake, 
G.~A., Qi, C., Dullemond, C.~P., \& Wilner, D.~J.\ 2008, \apjl, 675,
L109 

\bibitem[Brown et al.(2009)]{brown09} Brown, J.~M., Blake, 
G.~A., Qi, C., et al.\ 2009, \apj, 704, 496

\bibitem[Brown et al.(2012)]{brown2012} Brown, J.~M., Rosenfeld, 
K.~A., Andrews, S.~M., Wilner, D.~J., 
\& van Dishoeck, E.~F.\ 2012, \apjl, 758, L30

\bibitem[Calvet et al.(2002)]{calvet2002} Calvet, N., D'Alessio, 
P., Hartmann, L., et al.\ 2002, Apj, 568, 1008

\bibitem[Calvet et al.(2005)]{calvet2005} Calvet, N., D'Alessio, 
P., Watson, D.~M., et al.\ 2005, \apjl, 630, L185


\bibitem[Casassus et al.(2013)]{casassus13} Casassus, S., van der 
Plas, G., M, S.~P., et al.\ 2013, \nat, 493, 191 

\bibitem[de Juan Ovelar et 
al.(2013)]{juanovelar2013} de Juan Ovelar, M., Min, M., Dominik, C., et al.\ 2013, \aap, 560, A111

\bibitem[Draine(2006)]{draine2006} Draine, B.~T.\ 2006, \apj, 636, 
1114

\bibitem[Dorschner et 
al.(1995)]{dorschner1995} Dorschner, J., Begemann, B., Henning, T., Jaeger, C., \& Mutschke, H.\ 1995, \aap, 300, 503

\bibitem[{{Espaillat} {et~al.}(2010){Espaillat}, {D'Alessio}, {Hern{\'a}ndez},
  {Nagel}, {Luhman}, {Watson}, {Calvet}, {Muzerolle}, \&
  {McClure}}]{espaillat10}
{Espaillat}, C., {D'Alessio}, P., {Hern{\'a}ndez}, J., {et~al.} 2010, \apj,
  717, 441

\bibitem[Espaillat et al.(2011)]{espaillat2011} Espaillat, C., 
Furlan, E., D'Alessio, P., et al.\ 2011, \apj, 728, 4

\bibitem[Espaillat et al.(2012)]{espaillat12} Espaillat, C., 
Ingleby, L., Hern{\'a}ndez, J., et al.\ 2012, \apj, 747, 103

\bibitem[Finkbeiner et al.(1999)]{Finkbeiner1999} Finkbeiner, D.~P., 
Davis, M., \& Schlegel, D.~J.\ 1999, ApJ, 524, 867

\bibitem[Fukagawa et al.(2013)]{fukagawa2013} Fukagawa, M., 
Tsukagoshi, T., Momose, M., et al.\ 2013, arXiv:1309.7400 

\bibitem[Furlan et al.(2006)]{furlan2006} Furlan, E., Hartmann, 
L., Calvet, N., et al.\ 2006, ApJS, 165, 568

\bibitem[Garufi et 
al.(2013)]{Garufi2013} Garufi, A., Quanz, S.~P., Avenhaus, H., et al.\ 2013, \aap, 560, A105

\bibitem[Guilloteau et 
al.(2011)]{Guilloteau2011} Guilloteau, S., Dutrey, A., Pi{\'e}tu, V., \& Boehler, Y.\ 2011, A\&A, 529, A105 

\bibitem[Henning et 
al.(1995)]{henning1995} Henning, T., Michel, B., \& Stognienko, R.\ 1995, \planss, 43, 1333 

\bibitem[Henning 
\& Stognienko(1996)]{henning1996} Henning, T., \& Stognienko, R.\ 1996, \aap, 311, 291

\bibitem[Isella et 
al.(2007)]{isella2007} Isella, A., Testi, L., Natta, A., et al.\ 2007, \aap, 469, 213 


\bibitem[Isella et al.(2010a)]{isella2010a} Isella, A., Carpenter, 
J.~M., \& Sargent, A.~I.\ 2010a, ApJ, 714, 1746 

\bibitem[Isella et al.(2012)]{isella2012} Isella, A., P{\'e}rez, 
L.~M., \& Carpenter, J.~M.\ 2012, \apj, 747, 136

\bibitem[Isella et al.(2013)]{isella2013} Isella, A., P{\'e}rez, 
L.~M., Carpenter, J.~M., et al.\ 2013, \apj, 775, 30

\bibitem[Jaeger et 
al.(1994)]{jaeger1994} Jaeger, C., Mutschke, H., Begemann, B., Dorschner, J., \& Henning, T.\ 1994, \aap, 292, 641

\bibitem[Klahr 
\& Henning(1997)]{Klahr1997} Klahr, H.~H., \& Henning, T.\ 1997, Icarus, 128, 213

\bibitem[Kelly(2007)]{kelly2007} Kelly, B.~C.\ 2007, \apj, 665, 
1489

\bibitem[{{Lin} \& {Papaloizou}(1979)}]{lin79}
{Lin}, D.~N.~C. \& {Papaloizou}, J. 1979, \mnras, 186, 799

\bibitem[Lommen et 
al.(2007)]{lommen2007} Lommen, D., Wright, C.~M., Maddison, S.~T., et al.\ 2007, \aap, 462, 211 

\bibitem[Lommen et 
al.(2009)]{lommen2009} Lommen, D., Maddison, S.~T., Wright, C.~M., et al.\ 2009, \aap, 495, 869 

\bibitem[Lommen et 
al.(2010)]{lommen2010} Lommen, D.~J.~P., van Dishoeck, E.~F., Wright, C.~M., et al.\ 2010, \aap, 515, A77

\bibitem[Lynden-Bell 
\& Pringle(1974)]{Lyden-Bell1974} Lynden-Bell, D., \& Pringle, J.~E.\ 1974, MNRAS, 168, 60

\bibitem[Lyra 
\& Lin(2013)]{lyra2013} Lyra, W., \& Lin, M.-K.\ 2013, \apj, 775, 17 

\bibitem[Mathews et al.(2012)]{mathews2012} Mathews, G.~S., 
Williams, J.~P., \& M{\'e}nard, F.\ 2012, \apj, 753, 59

\bibitem[McCabe et al.(2006)]{mcabe2006} McCabe, C., Ghez, A.~M., 
Prato, L., et al.\ 2006, \apj, 636, 932

\bibitem[{{Masset}(2000)}]{masset00}
{Masset}, F. 2000,A\&AS, 141, 165

\bibitem[Natta 
\& Testi(2004)]{natta2004a} Natta, A., \& Testi, L.\ 2004a, Star
Formation in the Interstellar Medium: In Honor of David Hollenbach,
323, 279 

\bibitem[Natta et al.(2007)]{natta2007} Natta, A., Testi, L., 
Calvet, N., et al.\ 2007, Protostars and Planets V, 767

\bibitem[Olofsson et 
al.(2013)]{olofsson2013} Olofsson, J., Benisty, M., Le Bouquin, J.-B., et al.\ 2013, \aap, 552, A4 

\bibitem[Owen 
\& Clarke(2012)]{owen2012} Owen, J.~E., \& Clarke, C.~J.\ 2012, \mnras, 426, L96 

\bibitem[Paardekooper 
\& Mellema(2004)]{paardekooper04} Paardekooper, S.-J., \& Mellema, G.\ 2004, \aap, 425, L9

\bibitem[P{\'e}rez et al.(2012)]{perez2012} P{\'e}rez, L.~M., 
Carpenter, J.~M., Chandler, C.~J., et al.\ 2012, ApJL, 760, L17

\bibitem[P{\'e}rez et al.(2014)]{perez2014} P{\'e}rez, L.~M., 
Isella, A., Carpenter, J.~M., \& Chandler, C.~J.\ 2014, arXiv:1402.0832 

\bibitem[{{Pi{\'e}tu} {et~al.}(2006){Pi{\'e}tu}, {Dutrey}, {Guilloteau},
  {Chapillon}, \& {Pety}}]{pietu06}
{Pi{\'e}tu}, V., {Dutrey}, A., {Guilloteau}, S., {Chapillon}, E., \& {Pety}, J.
  2006, \aap, 460, L43

\bibitem[Pinilla et 
al.(2012b)]{pinilla2012} Pinilla, P., Birnstiel, T., Ricci, L., et al.\ 2012b, \aap, 538, A114

\bibitem[Pinilla et 
al.(2012a)]{pinilla12a} Pinilla, P., Benisty, M., \& Birnstiel, T.\
2012a, \aap, 545, A81 

\bibitem[Pinilla et 
al.(2013)]{pinilla2013} Pinilla, P., Birnstiel, T., Benisty, M., et al.\ 2013, \aap, 554, A95

\bibitem[Qi et al.(2011)]{qi2011} Qi, C., D'Alessio, P., 
{\"O}berg, K.~I., et al.\ 2011, \apj, 740, 84

\bibitem[Quanz et al.(2013)]{quanz2013} Quanz, S.~P., Avenhaus, 
H., Buenzli, E., et al.\ 2013, \apjl, 766, L2 

\bibitem[Reg{\'a}ly et al.(2012)]{Regaly2012} Reg{\'a}ly, Z., 
Juh{\'a}sz, A., S{\'a}ndor, Z., 
\& Dullemond, C.~P.\ 2012, MNRAS, 419, 1701 

\bibitem[{{Rice} {et~al.}(2006){Rice}, {Armitage}, {Wood}, \&
  {Lodato}}]{rice06}
{Rice}, W.~K.~M., {Armitage}, P.~J., {Wood}, K., \& {Lodato}, G. 2006, \mnras,
  373, 1619

\bibitem[Ricci et 
al.(2010a)]{Ricci2010a} Ricci, L., Testi, L., Natta, A., Neri, R., Cabrit, S., \& Herczeg, G.~J.\ 2010a, A\&A, 512, A15 

\bibitem[Ricci et 
al.(2010b)]{Ricci2010b} Ricci, L., Testi, L., Natta, A., \& Brooks, K.~J.\ 2010b, A\&A, 521, A66

\bibitem[Ricci et 
al.(2011)]{Ricci2011} Ricci, L., Mann, R.~K., Testi, L., Williams, J.~P., Isella, A., Robberto, M., Natta, A., \& Brooks, K.~J.\ 2011, A\&A, 525, A81

\bibitem[Ricci et al. (2012)]{Ricci2012} Ricci, L., Testi, L., Natta,
  A., Scholz, A. \& de Gregorio-Monsalvo, I., ApJ, 761, L20

\bibitem[Ricci et 
al.(2012)]{Ricci2012b} Ricci, L., Trotta, F., Testi, L., et al.\ 2012, \aap, 540, A6 

\bibitem[Rodmann et 
al.(2006)]{rodmann2006} Rodmann, J., Henning, T., Chandler, C.~J., Mundy, L.~G., \& Wilner, D.~J.\ 2006, \aap, 446, 211

\bibitem[Rosotti et al.(2013)]{rosotti2013} Rosotti, G.~P., 
Ercolano, B., Owen, J.~E., \& Armitage, P.~J.\ 2013, MNRAS, 430, 1392


\bibitem[Semenov et 
al.(2003)]{Semenov2003} Semenov, D., Henning, T., Helling, C., Ilgner, M., \& Sedlmayr, E.\ 2003, A\&A, 410, 611 

\bibitem[Stognienko et al.(1996)]{stognienko1996} Stognienko, R., 
Henning, T., 
\& Ossenkopf, V.\ 1996, IAU Colloq.~150: Physics, Chemistry, and Dynamics of Interplanetary Dust, 104, 427 


\bibitem[Testi et al.(2001)]{testi2001} Testi, L., Natta, A., 
Shepherd, D.~S., \& Wilner, D.~J.\ 2001, \apj, 554, 1087 

\bibitem[Testi et 
al.(2003)]{Testi2003} Testi, L., Natta, A., Shepherd, D.~S., \&
Wilner, D.~J.\ 2003, \aap, 403, 323 

\bibitem[Testi et al.(2014)]{testi2014} Testi, L., Birnstiel, T., 
Ricci, L., et al.\ 2014, arXiv:1402.1354


\bibitem[Trotta et 
al.(2013)]{trotta2013} Trotta, F., Testi, L., Natta, A., Isella, A., \& Ricci, L.\ 2013, \aap, 558, A64

\bibitem[Ubach et al.(2012)]{ubach2012} Ubach, C., Maddison, 
S.~T., Wright, C.~M., et al.\ 2012, \mnras, 425, 3137

\bibitem[van der Marel et al.(2013)]{vandermarel2013} van der Marel, 
N., van Dishoeck, E.~F., Bruderer, S., et al.\ 2013, Science, 340,
1199

\bibitem[White 
\& Ghez(2001)]{white2001} White, R.~J., \& Ghez, A.~M.\ 2001, \apj, 556, 265 

\bibitem[{{Williams} \& {Cieza}(2011)}]{williams11}
{Williams}, J.~P. \& {Cieza}, L.~A. 2011, ARA\&A, 49, 67

\bibitem[Wilner et al.(2003)]{wilner2003} Wilner, D.~J., Bourke, 
T.~L., Wright, C.~M., et al.\ 2003, \apj, 596, 597

\bibitem[Windmark et 
al.(2012)]{windmark2012} Windmark, F., Birnstiel, T., G{\"u}ttler, C., et al.\ 2012, \aap, 540, A73 
 
\bibitem[Zhu et al.(2012)]{zhu2012} Zhu, Z., Nelson, R.~P., 
Dong, R., Espaillat, C., \& Hartmann, L.\ 2012, \apj, 755, 6


\bibitem[Zhu et al.(2013)]{zhu2013} Zhu, Z., Stone, J.~M., 
Rafikov, R.~R., \& Bai, X.\ 2013, arXiv:1308.0648
\end{thebibliography}

\end{document}